\documentclass[12pt, ChapStyle1, oneside]{./styles/pfe}
\usepackage[utf8]{inputenc}
\usepackage[T1]{fontenc}

\usepackage{longtable}
\usepackage[ruled,vlined,french,titlenumbered]{algorithm2e}
\usepackage{mathrsfs, amsmath, amssymb, bbm}
\usepackage[colorlinks=true,linkcolor=black, citecolor=black, pdfhighlight =/O]{hyperref}
\usepackage[usenames]{color}
\usepackage{pstricks, pst-node, pst-tree}
\usepackage{graphicx,graphics,epsfig}
\usepackage{floatflt, multirow, array, subfigure, hhline, enumerate, comment, url, pifont}
\usepackage[active]{rotating}
\usepackage{multicol}
\usepackage{graphicx}
\usepackage{xcolor}

\begin{document}

\thispagestyle{empty}
  \vspace*{-3.2cm}

\begin{center}
\textsc{république tunisienne} \\
\textsc{ministère de l'enseignement supérieur},\\
\textsc{de la recherche scientifique et de la technologie} \\
\textsc{université de tunis el manar} \\
\end{center}

\begin{center}
\begin{figure}[htbp]
    \centering
       \includegraphics[width=5cm,height=3cm]{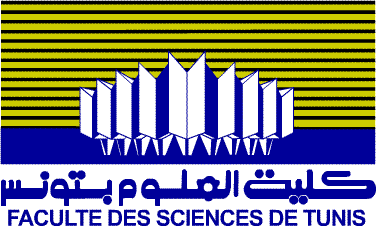}
\end{figure}
\bigskip
\textsc{faculté des sciences de tunis} \\
\textsc{département des sciences de l'informatique} \\
\end{center}

 \bigskip
\begin{center}
\Large{\textbf{ \textsc{mémoire de mastère en informatique} \\}}

\small { présenté en vue de l'obtention du \\
\textbf{Diplôme de Mastère en Informatique}\\}
\end{center}

\begin{center}

\bigskip

 \LARGE \textbf{Détection et extraction des particules d'intérêts dans une image biologique 3D}

\bigskip
\end{center}

\begin{center}
 \large Présenté par :  \large{\textbf{Mariam  \textsc{El Oussini}}}\\
\end{center}

 \normalsize
\bigskip\bigskip
\bigskip\bigskip
\bigskip\bigskip
\bigskip\bigskip

\begin{flushleft}
\large Soutenu le 19 septembre 2019 devant le jury d'examen composé de :

\begin{tabular}{ll}
Président  : & \textbf{M.  Faouzi \textsc{Moussa}} (Professeur, FST)\\
Rapporteur : & \textbf{M.  Atef \textsc{Hamouda}} (Ma\^{i}tre assistant, FST) \\
Directeur de mémoire : & \textbf{M. Mohamed \textsc{Naouai}} (Ma\^{i}tre assistant, FST)\\

\end{tabular}
\end{flushleft}

\bigskip\bigskip
\bigskip\bigskip
\bigskip\bigskip

\begin{center}

 \large Année universitaire : 2018-2019

\end{center}

\cleardoublepage
\pagenumbering{roman}
\setcounter{page}{1}

\newpage
\thispagestyle{empty}
  \vspace*{-3cm}

\begin{center} \Large Résumé \end{center}

\PARstart{L}{a} segmentation cellulaire est un domaine de recherche en pleine expansion. En effet, les méthodes classiques de segmentation ne suffiront pas pour segmenter ce type d'images. Dans ce manuscrit, nous présentons une nouvelle méthode permettant la segmentation des ribosomes. Une phase de prétraitement précède la segmentation et une autre phase de post-traitement qui la suit. La principale originalité de notre proposition est la qualité de l'objet 3D segmenté qu'elle fournit. Nous menons une étude expérimentale qui vise à prouver l'efficacité de notre approche.

\paragraph*{Mots clés~:} Extraction des ribosomes, Segmentation cellulaire, Tomographie électronique, Tomogramme. 
\vspace{0.5cm}

\begin{center} \Large Abstract \end{center}

\PARstart{C}{ells} segmentation shows rapid growth in biology. Indeed, using the classical segmentation methods only is not enough to segment this type of images. In this manuscript, we will present a new method of ribosomes segmentation. A pre-treatment phase will precedes the segmentation process and after that a post-processing will proceed. 

\paragraph*{Key words:} Ribosome extraction, Cells segmentation, Electron tomography, Tomogram.\newpage

\thispagestyle{empty}

\begin {center}
\huge{\textbf{Dédicaces}}
\end {center}
\textit{En témoignage de ma gratitude, de mon amour et de ma grande reconnaissance, je dédie ce travail à :} \newline

\textit{Mon père Lotfi, pour son soutien et amour,  je vous remercie d'être toujours là pour moi,} \newline

\textit{Ma mère Monia, pour sa confiance en moi et son amour,} \newline

\textit{Mes soeurs qui m'ont énormément encouragé,} \newline

\textit{Toute ma famille,} \newline

\textit{Mes amis,} \newline

\textit{Mes enseignants à la Faculté des Sciences de Tunis.} \newline

\Remerciements{Remerciements} \label{Remerciements}

\noindent Un travail qui est dû d'un effort humain est toujours le fruit d'une collaboration. \newline\newline
 Que tous ceux qui ont pris une part dans la réalisation de ce travail puissent trouver dans ces lignes l'expression de notre profonde gratitude.\newline\newline
Nous commençons particulièrement par mon encadrant, M. Mohamed \textsc{Naouai}, qui a bien voulu accepter de diriger ce travail. Nous lui remercions énormément.\newline\newline
Le m\^{e}me sentiment de reconnaissance est aussi exprimé à M. Hmida \textsc{Rojbani}, pour sa disponibilité, ses directions et son aide tout au long de mes travaux de recherches. \newline\newline
Nos remerciements vont aussi à l'endroit de tous nos chers professeurs qui ont fait de ce que nous sommes aujourd'hui. \newline\newline

\tableofcontents
\listoffigures %
\mainmatter%

\pagenumbering{arabic}
\setcounter{page}{1}
  \Introduction{Introduction générale} \label{Introdution} %

\PARstart{L}{a} tomographie électronique, qui est une technologie d'imagerie tridimensionnelle utilisée généralement en biologie, est une des techniques utilisées pour visualiser en détail des structures de cellules végétales ou animales, bactéries, virus, protéine, petite molécule ou bien atomes.
Cette technique permet d'estimer la densité 3D d'un objet à partir d'une série d'images 2D capturées par microscope électronique sous différents angles d'un objet quelconque. 
Le processus de la tomographie électronique sera comme suit : tout d'abord l'acquisition de série d'images 2D, l'alignement de cette série d'images et par la suite la reconstruction 3D. \\ Certains techniques de filtrage et de segmentation existent pour pouvoir analyser les images biologiques. Notre travail consiste à améliorer et segmenter les images biologiques obtenues à partir du tomogramme reconstruit par un technique que nous avons développé. \\ 
Dans ce manuscrit, nous aurons quatre chapitres  :  dans le premier chapitre, nous allons décrire brièvement les différentes étapes de la tomographie électronique et la méthode de reconstruction utilisée pour produire les images que nous allons traiter puisqu'ils sont essentiels pour comprendre la méthode de segmentation que nous allons proposer par la suite. Dans le deuxième chapitre, nous allons décrire quelques méthodes de segmentation d'images biologiques. Ensuite dans le troisième chapitre, nous allons décrire notre approche et enfin le dernier chapitre comporte l'évaluation de l'approche.

 \chapter{ Contexte et problématique} \label{ch1} 
\section*{Introduction}
Dans ce premier chapitre, nous allons tout d'abord introduire les concepts de base des différents types de microscopes, par la suite nous allons décrire les différentes étapes de la tomographie électronique tels que l'acquisition, l'alignement, la reconstruction et enfin la segmentation, en expliquant ses différentes phases. Et enfin dans la dernière partie de ce chapitre nous présenterons la problématique.
\section{Historique}
En 1929, la microscopie électronique est apparu grâce aux travaux de Louis de Broglie~\cite{louisb} qui a découvert qu'il est possible de traiter les électrons comme une onde, et les utiliser par la suite sous forme de faisceau lumineux dans un microscope. C'est à partir de ces travaux que Ernst Ruska a con\c{c}u les premières lentilles électromagnétiques, qui ont été utilisées au premier microscope électronique en 1931. Après cette découverte, Ernst Ruska a reçu le prix Nobel en 1986~\cite{metwc}.
Ensuite, la théorie du contraste de diffraction électronique a été développée par certains chercheurs de l'Université de Cambridge qui a pour but de déterminer les structures de cristal des matériaux cristallins~\cite{diffcont} à partir de 1970, de nombreux Microscopes électroniques ont été conçus~\cite{metwc} et progressivement la microscopie et l'imagerie biologique occupent une place très importante. 
   

\section{La microscopie}  
\begin{figure}[!htbp]
  \centering
\includegraphics[scale=0.5]{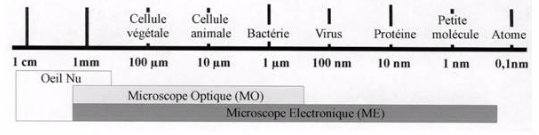} \\
\caption{L'utilisation des microscopes}\label{mics} 
\end{figure}

L'observation des détails extrêmement petits d'un objet animal, plante ou roche est observée à l'aide d'un microscope à transmission photonique ou électronique (\textit{cf. figure}  \ref{mics}).
La différence entre ses deux types de microscopes réside dans la nature des lentilles (en électromagnétiques ou en verre), la nature de la source d'énergie (électrons ou photons), et le mode d'observation qui est effectué à l'aide d'un écran fluorescent ou par l'oeil~\cite{mic}.

\subsection{Microscope optique}
\begin{figure}[!htbp]
  \centering
\includegraphics[scale=1.5]{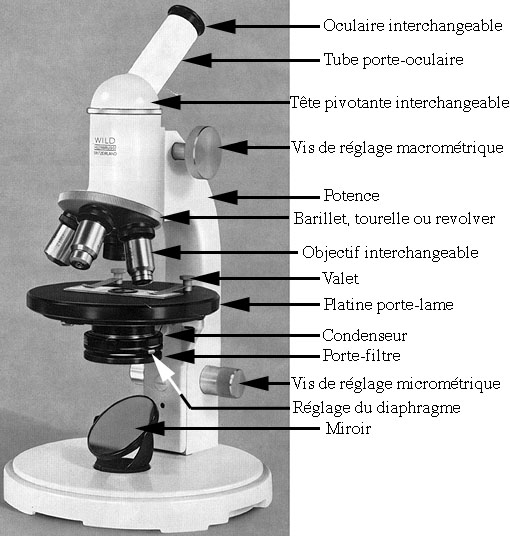}\\
\caption{La microscope optique}\label{opt} 
\end{figure}

Les objets très fins peuvent être visualisés à l'aide d'un microscope optique. L'objet monté dans une goutte d'eau ou bien dans un liquide coloré sur une lame porte-objet. Ce dernier recouvert par une lamelle fragile (couvre-objet) ou des coupes fixées et colorées. 
Le processus d'acquisition sera comme suit. La lumière traverse l'objet et par la suite remonte dans les lentilles de verre et finalement l'objet s'agrandit (\textit{cf. figure}\ref{opt}).

L'observation au microscope se fait comme suit : tout d'abord, l'objet à observer est placé sur la platine et centré pour que la lumière puisse traverser le tube optique en donnant un rond lumineux dans l'oculaire. Ensuite, l'objet est placé dans l'axe du tube optique et il suffit de regarder l'oculaire en modifiant à l'aide de la vis macrométrique l'image pour qu'elle soit plus nette. La préparation (l'objet à observer) est déplacée soigneusement jusqu'à trouver l'objet recherché. Une fois trouvé, placer la zone à agrandir au centre de la platine, ensuite en tournant le barillet pour changer d'objectif, sans modifier le réglage précédent~\cite{opmic}.

\subsection{Microscope électronique}
\begin{figure}[!htbp]
  \centering
\includegraphics[scale=0.5]{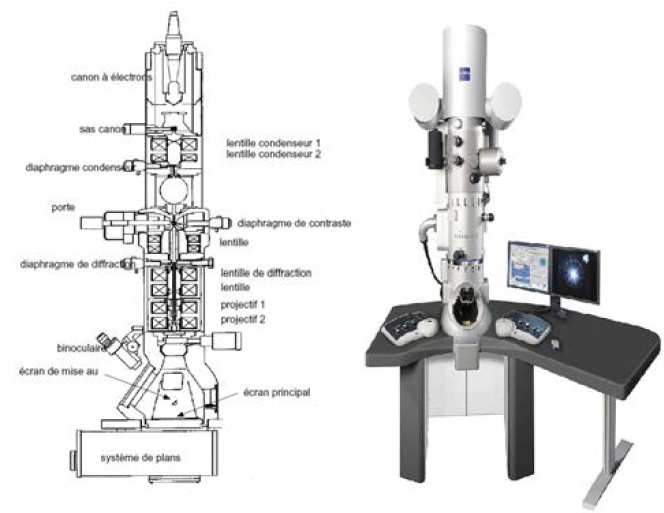} \\
\caption{Microscope électronique en transmission~\cite{mic}}\label{mic}
\end{figure}

Il existe plusieurs types de microscopes électroniques parmi lesquels le microscope électronique à balayage (MEB), microscope électronique en transmission (MET) (\textit{cf. figure}  \ref{mic}) et microscope à êépiflurorescence.\\
Il est à noter que l'analyse des échantillons biologiques dépend du type utilisé. Par exemple pour un MEB, on mesure les faisceaux électroniques réfléchis à la surface de l'échantillon et une image sera construite pixel par pixel tout en mesurant les électrons secondaires ainsi rétrodiffusés. Et pour un MET les faisceaux électroniques passent à travers l'échantillon et ensuite ces derniers seront amplifiés par un ensemble de lentilles électromagnétiques dans le but de produire une image~\cite{a3}.

\subsubsection{Interaction électrons-matière}

\begin{figure}[!htbp]
  \centering
\includegraphics[scale=0.7]{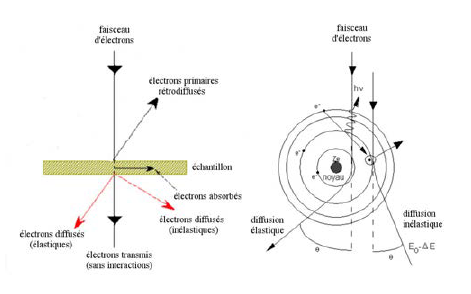} \\
\caption{Diffusion élastique et inélastique entre un électron incident de haute
énergie et un atome~\cite{mic}}\label{elecmat}
\end{figure}

Vu que la masse est faible, la déviation de l'électron par le noyau ou les électrons de l'atome est possible. Dans les techniques d'imagerie MET, le processus principal est la diffusion de particule par des interactions électrostatique ou interactions de Coulomb. Il est possible de traiter l'électron avec deux manières différentes : soit comme étant une particule, soit comme étant une onde durant la diffusion d'électrons~\cite{metwc}.
Il existe principalement deux formes de diffusion. L'une avec perte d'énergie et l'autre sans perte, respectivement élastique et inélastique \ref{elecmat}. Durant l'analyse des échantillons, les deux formes sont utiles. Mais, il est possible que le spécimen soit endommagé, vu que ses formes ne sont pas convenables pour les matières biologiques.  Quel que soit le type de la forme élastique ou inélastique, elles diminuent l'efficacité de l'analyse d'échantillons pour le MET~\cite{elecmat1}. \\
Pour résoudre ce problème, il faut que l'échantillon soit préparé par une méthode de fixation. En plus, il faut garder les propriétés des échantillons à étudier, en utilisant ou bien développant une des techniques de préparation~\cite{elecmat2}. \\
Pour une image à haute résolution, un échantillon biologique avec une épaisseur de 100 nm est suffisant~\cite{elecmat3}.

\subsubsection{Microscope électronique à transmission (MET)}
\begin{figure}[!htbp]
  \centering
\includegraphics[scale=0.5]{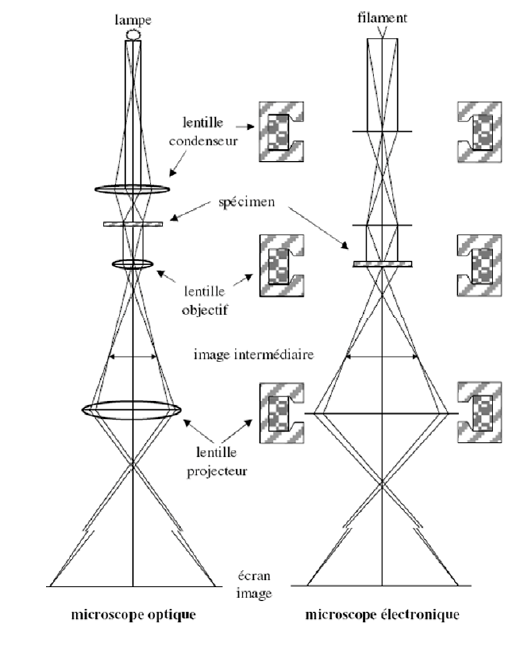} \\
\caption{Microscope optique et électronique}\label{micopelec}
\end{figure}

Dans le domaine de la biologie, l'outil d'acquisition le plus utilisé est le MET. Grâce à ce microscope, l'acquisition des images 2D s'effectuera. Ce microscope permet d'étudier des coupes fines de cellules aussi bien à réaliser des études structurales sur des complexes protéiques isolés ou des virus à haute résolution. Grâce à ce type de microscope et après traitement d'images, l'obtention d'un modèle 3D d'objets biologiques comme les ribosomes est possible~\cite{a2}.
Un MET fonctionne partiellement comme un microscope optique, la seule différence le MET utilise les électrons comme source de lumière au lieu des photons aussi bien que les rayonnements traversent l'échantillon et par la suite les images se projettent sur un écran phosphorescent à la place de la rétine observatrice (\textit{cf. figure} \ref{micopelec}).
 
\begin{figure}[!htbp]
  \centering
\includegraphics[scale=0.7]{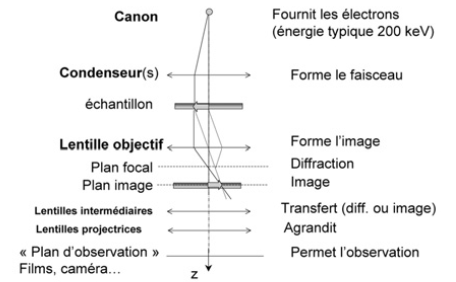} \\
\caption{Principe d'un MET~\cite{a3}}\label{tem}
\end{figure}

Conceptuellement un MET existe sur deux formes. Les TEM (\textit{Transmission Electron Microscope}) comme dans le cas d'un microscope de paillasse, le faisceau est une plane incidente sur l'échantillon et les STEM semblable au microscope optiques confocaux  (\textit{Scanning Transmission Electron Microscope}). Le plus important à retenir c'est que les microscopes modernes fonctionnent dans les deux modes.  \\
Le mode du fonctionnement d'un TEM est schématisé dans la figure \ref{tem}. Pour régler la taille du faisceau et l'angle d'incidence, ce dernier passe par une série de lentilles condenseur. Une première image sera produite par la lentille objective, une fois le faisceau atteint l'échantillon. La résolution de l'image dépend de la qualité de la lentille objective. En plus cette dernière n'agrandit pas (= x 10), les lentilles protectrices prennent en charge cette tâche.  \\
à cette étape, si le plan objet de l'image co\"{i}ncide avec le plan image de la lentille objectif, la lentille intermédiaire formera une deuxième image. Sinon si le plan objet co\"{i}ncide avec le plan focal de la lentille objectif, la lentille intermédiaire formera une image du cliché de diffraction. \\
L'un des avantages des MET, l'obtention de l'image ou le cliché de diffraction de la même zone est simple, en modifiant la valeur de la distance focale de la lentille intermédiaire. En pratique, un bouton modifie le courant des bobines intermédiaires.\\
Pour finir, l'image ou le cliché de diffraction formé sera donc agrandi et projeté sur le détecteur grâce aux lentilles protectrices. Le plan d'observation sur lequel l'image sera projetée peut être une plaque photo, un écran fluorescent ou bien un scintillateur couplé à une caméra CCD.

\begin{figure}[!htbp]
  \centering
\includegraphics[scale=0.7]{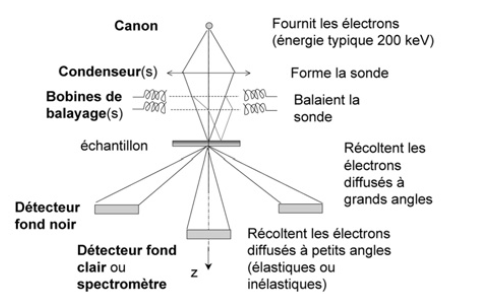} \\
\caption{Principe d'un STEM~\cite{a3}}\label{stem}
\end{figure}

\newpage
Comme c'est déjà mentionné dans la section précédente. En mode TEM, par un faisceau larde, l'échantillon sera illuminé et une résolution spatiale se fait par le détecteur. Contrairement en mode STEM (\textit{cf. figure}  \ref{stem}), on mesure les signaux d'intérêt en chaque point d'un balayage conclu à partir d'un faisceau focalisé sur l'échantillon. La plupart des cas, ces signaux sont mesurés sur un détecteur non résolu spatialement. Donc l'image est formée d'une manière séquentielle, et l'intensité du signal mesuré sera attribuée pour chaque point du balayage. 

\newpage

Les restrictions qu'un MET dispose sont les suivants : \\
\begin{itemize}
\item Le bruit dans les images acquises est élevé. Ce qui mène à diminuer la qualité de l'image acquise. Par exemple, le rapport: signal sur bruit, de la figure \ref{cryofx}, est 5,16 dB.

\begin{figure}[!htbp]
  \centering
\includegraphics[scale=0.7]{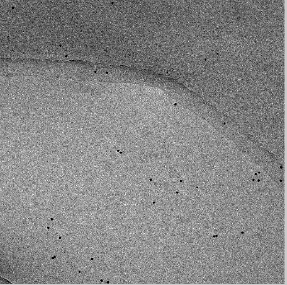} \\
\caption{Images obtenue par cryo-fixation d'une tranche de cellule eucaryote}\label{cryofx}
\end{figure}

\item Un autre problème en MET, l'épaisseur du spécimen doit être mince. Plus que l'échantillon à préparer est mince plus que sa préparation est encore plus difficile aussi bien que les spécimens minces ont un contraste faible.
\item Une autre restriction, d'un MET, lors de la phase de reconstruction, nous aurons des projections manquantes. Cette perte d'information s'explique sur le fait que le spécimen ne peut pas être basculé à 360$^{\circ}$. Généralement, l'intervalle d'inclinaison du spécimen est de -70$^{\circ}$ à +70$^{\circ}$.
\end{itemize}

Les restrictions que nous venons de les décrire au-dessus, nous allons maintenant préciser, respectivement, les causes de ces derniers  : 
\begin{itemize}
\item La préparation du spécimen est la tâche la plus importante lors de utilisation d'un MET, vu que les résultats dépendent d'une bonne fixation. La fixation chimique et cryogénique (\textit{cf. figure}  \ref{spec}) est l'un des techniques les plus utilisés. La différence entre la fixation chimique et cryogénique, en utilisant le premier technique nous risquons d'avoir une information reconstruite inexacte, vu que cette fixation peut déformer le spécimen par contre le deuxième technique préserve sa structure interne grâce à la congélation à une température de l'azote liquide~\cite{spec1}. Cette méthode pose des contraintes au niveau de l'énergie du faisceau d'électrons. Une fois, le faisceau chauffe le spécimen, son énergie devient élevée et le spécimen devient décongelé ainsi évaporé. Ce que nous venons de citer n'est pas la seule cause qui nécessite que l'énergie de faisceau ne soit pas élevée. Mais vu que les électrons sont un type particulier de radiation ionique ce qui fait qu'ils peuvent déformer le spécimen. Dans le but d'éviter cette situation, il faut réduire au maximum la dose totale d'électrons dont un spécimen pourra le recevoir. En diminuant cette énergie du faisceau, le bruit dans les images acquises augmente.

\begin{figure}[!htbp]
  \centering
\includegraphics[scale=0.5]{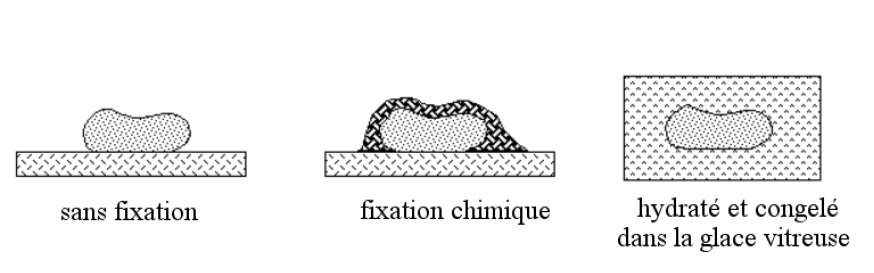} \\
\caption{Méthodes de fixation }\label{spec}
\end{figure}

\item D'abord pour tous les microscopes électroniques en transmission, la préparation du spécimen doit être dans un disque ou une grille de 3 mm de diamètre sous forme de lame mince~\cite{a3} aussi bien que l'épaisseur doit être plus mince que 100 nm. Sachant que, dans un délai raisonnable, une quantité d'électrons passe à travers l'échantillon pour obtenir des informations interprétables. Le spécimen doit être suffisamment transparent aux électrons~\cite{metwc}.

\item Si l'angle d'inclinaison est supérieur à 70$^{\circ}$. Le spécimen devient opaque pour les électrons. Autrement dit, le spécimen ne se basculera pas à 360$^{\circ}$ degrés. Généralement, ce dernier pourrait être incliné à -70$^{\circ}$ à +70$^{\circ}$. L'apparition d'une zone aveugle dans l'espace de fourrier (\textit{cf. figure}  \ref{incl})  est due à cause de problèmes du prisme triangulaire manquant dans l'espace d'acquisition~\cite{axspec}.

\begin{figure}[!htbp]
  \centering
\includegraphics[scale=0.5]{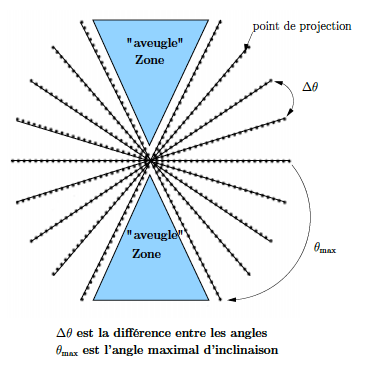} \\
\caption{L'échantillonnage des données dans l'espace de Fourier}\label{incl}
\end{figure}

\newpage
Cette zone manquante que nous venons de mentionner aura un impact aux résultats obtenus durant la phase de reconstruction. Nous aurons une perte d'information ce qui explique la perte de résolution et des artefacts d'épandage en étoile due aux effets d'interpolation.  (\textit{cf. figure}  \ref{trans133}).

\end{itemize}

\begin{figure}[!htbp]
  \centering
\includegraphics[scale=0.7]{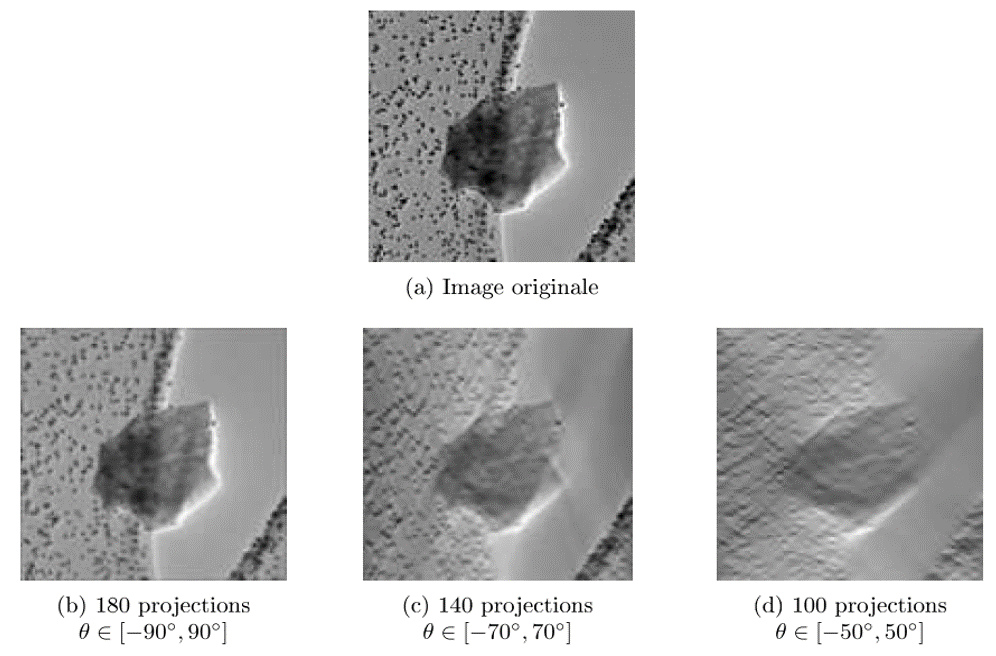} \\
\caption{L'influence des angles manquants sur les objets reconstruits}\label{trans133}
\end{figure}

    \newpage
\section{Imagerie tridimensionnelle en biologie } 
\begin{figure}[!htbp]
  \centering
\includegraphics[scale=0.6]{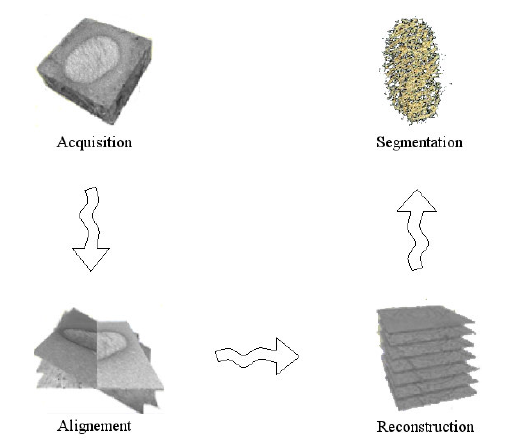} \\
\caption{Etapes du processus de reconstruction 3D tomorgraphie}\label{te}
\end{figure}

 En nous basant sur une série d'images de projections 2D et en appliquant la tomographie d'électrons, nous pouvons reconstruire une structure 3D de l'échantillon biologique. 
Cette technique appelée tomographie électronique est composée de quatre étapes (\textit{cf. figure} \ref{te}). Premièrement, l'acquisition d'une série d'images de projections 2D sous plusieurs angles d'inclinaison par un microscope électronique en transmission (MET) aussi bien qu'elles sont faiblement contrastée ce qui mène à une confusion entre la zone d'intérêt et l'arrière-plan. Pour cette raison, elles nécessitent un prétraitement pour améliorer leurs qualités~\cite{a1}. Deuxièmement, cette série d'images obtenues à l'étape précédente doit être alignée sur un axe d'inclinaison commun. Dans le but d'éliminer les décalages et les rotations relatives aux images successives. Troisièmement le calcul de la reconstruction, en appliquant un algorithme sur les images alignées à l'étape précédente. Et finalement dans la quatrième et dernière étape, à la suite de la reconstruction du volume 3D de l'échantillon biologique, la reconnaissance des structures qui le composent exige : des connaissances expertes, des méthodes de segmentation et classification aussi bien que des fonctions de décisions. à cette étape, l'objet 3D reconstruit durant les étapes précédentes doit être segmenté dans le but de faciliter les analyses de l'échantillon aux biologistes. \\
Cette technique de cryotomographie électronique permet de visualiser les objets biologiques avec une résolution nanométrique aussi bien qu'il assure l'estimation de la densité de l'objet reconstruit. 
En effet, les biologistes ne peuvent pas détecter les structures ou même les quantifier directement sans les phases de reconstruction ainsi l'extraction citée précédemment. 

\newpage
\subsection{Acquisition d'images et alignement d'une série d'images 2D}
Nous pouvons définir l'acquisition d'images comme étant une mesure spatiale d'une interaction entre l'onde et la matière. L'onde est émise par une source d'acquisition et re\c{c}u par un capteur. 
Notamment dans le cas de la TEM qui utilise comme rayonnement des électrons. Ce dernier a un système de lentilles magnétique qui permet de focaliser le rayon d'électrons sur un échantillon extrêmement mince. L'image que nous allons obtenir est transformée en image photonique. Par la suite, cette dernière sera enregistrée sur un film photographique ou depuis un capteur. \textit{charge-coupled device} (CCD).
Un capteur CCD, est un composant électronique photosensible qui permet de convertir un rayonnement électromagnétique en un signal électrique analogique. Par la suite, ce signal sera amplifié ainsi numérisé par un convertisseur analogique numérique pour obtenir une image numérique.
Cependant, nous obtenons à la fin de cette phase une série d'images de projection 2D à partir d'un MET. Cette série d'images est obtenue sur une plage d'inclinaison par un incrément angulaire de 1 ou 2 degrés, ensuite, alignée sur un axe d'inclinaison commun dans le but d'éliminer ces décalages.

\subsection{Reconstruction 3D}
En appliquant des méthodes analytiques ou algébriques sur une série d'images alignées de projection 2D, nous obtiendrons une structure 3D de l'objet. Les méthodes analytiques sont basées sur la transformée de Radon et les méthodes algébriques. Par exemple, nous traduisons le problème de la reconstruction en un ensemble d'équations linéaires :
 \begin{equation}\label{algeb}
  Ax=b
\end{equation}
o\`{u} $A$ est la matrice de projection, $x$ est l'objet d'origine et $b$ est l'ensemble des données de projection. En nous basant à une technique algébrique itérative, nous obtiendrons la solution au problème de reconstruction.

\subsection{Segmentation de tomogramme}
Une fois, nous obtenons l'objet 3D sous forme d'une image volumique en niveaux de gris appelé aussi tomogramme. Une segmentation sera effectuée sur les coupes 2D dans le but de faciliter les analyses ultérieures. \newline
L'objet à étudier est fondamentalement tridimensionnel. Dans notre cas, le tomogramme est une image tridimensionnelle (imagerie en coupes)~\cite{imgmed}. Nous rappelons aussi que  la segmentation assure la décomposition d'une image en des régions cohérentes et extraire les objets d'intérêt. Ce processus peut s'effectuer soit manuellement à l'aide d'un logiciel standard qui permet de visualiser des tomogrammes segmentés par exemple \textbf{UCSF Chimera}\footnote{\url{http://www.rbvi.ucsf.edu/chimera}}, ou bien par des méthodes de segmentation automatiques ou semi automatiques. Le choix de l'outil ou la méthode à utiliser dépend de la complexité d'étudier ces cellules, car manuellement la sélection est devenue plus coûteuse.
Généralement, la sélection de régions d'intérêts d'une image s'effectue en se basant sur des propriétés de pixels comme la couleur, la texture, les contours' etc. Ces propriétés ne sont pas suffisantes dans le cas de l'extraction à partir d'une image 3D (tomogramme). \newline
Dans la plupart des cas, les pixels de la région d'intérêt possèdent des intensités différentes par rapport à l'arrière-plan de l'image. La complexité de l'extraction se présente au niveau d'un faible rapport de signal-bruit aussi bien qu'il faut étudier la structure de l'objet à extraire.\newline
La performance des algorithmes de segmentation existante  dépend de l'objet à extraire. Il n'existe pas un algorithme standard qui assure l'extraction de tous les types de régions à sélectionner. 
En biologie, par exemple, la structure des cellules n'est pas les mêmes, ce qui fait, un algorithme peut-être plus efficace qu'un autre. Tout dépend de la structure de cellule à extraire. Dans le chapitre suivant, nous allons illustrer des algorithmes spécifiques développés pour la tomographie électronique.
\newline

Il est à noter que nous pouvons ajouter une étape de prétraitements avant la phase de segmentation, exactement juste après la reconstruction de l'objet 3D,  à fin de corriger les défauts liés à l'acquisition (réduire le bruit, ajuster le contraste, \textit{etc}).
Aussi bien que nous pouvons ajouter une étape de post-traitements dans le but de diminuer le taux d'erreurs et affiner les résultats obtenus.

\section{Problématique}
Nous avons déjà mentionné dans ce chapitre les limites d'un tomogramme tels que leur complexité, les artefacts dus par la zone aveugle d'acquisition, le bruit généré causé par le mouvement de l'objet lors de l'acquisition, \textit{etc}. 
Ces problèmes augmentent la complexité du processus de segmentation d'un tomogramme pour les images 3D.\newline
Avant de passer à la problématique, nous allons vous introduire l'intérêt et la technique d'acquisition de l'image que nous avons utilisée dans nos recherches.
\newline
Une cellule eucaryote est con\c{c}ue par plus d'un quart des ribosomes. Le r\^{o}le de ces derniers est d'effectuer la synthèse des protéines. Les images que nous allons utiliser dans nos recherches ont été capturées par un CMOS (Complementary Metal-Oxide Se- miconductor, détecteur d'électrons directes à haute sensibilité). Les images obtenues ont un contraste amélioré avec une taille 4096 $x$ 4096 pixels et une résolution de 3,4 \AA/pixel. \newline
En effet, notre travail est une continuité d'un travail de recherche, qui a été réalisé au sein de notre équipe de recherche~\cite{tomog3d}. Ils ont proposé deux méthodes : la première est une méthode d'alignement-reconstruction simultanée de tomogramme électronique qui permet d'optimiser le coût et la complexité de l'étape deux et trois de la tomographie électronique.  La deuxième est une méthode d'extraction de volume de ribosome qu'il traite également le problème de la segmentation des images 3D reconstruites avec une méthode de classification probabiliste. Vu que les images de cryotomographie révèlent des problèmes de bruit et de contraste. Deux méthodes de filtrage 3D ont été proposées comme prétraitement, la première est basée sur l'intégration fractionnaire et la deuxième est basée sur l'analyse multifractale.
\newline
Donc, à partir de ce travail nous allons aborder le problème de l'extraction de volume de ribosome différemment. Nous allons proposer une méthode de segmentation avec deux phases : la première est une phase de prétraitement pour l'amélioration d'images et la deuxième est un post-traitement pour la correction des erreurs aussi bien que l'affinement de la structure de l'objet reconstruit.
\section*{Conclusion}
Dans ce chapitre, nous avons expliqué le principe de la microscopie en transmission aussi bien que celle de la tomographie électronique. Bien évidemment,  nous avons décrit en détail les étapes de ce technique.

  \chapter{Segmentation d'images biologiques : état de l'art} \label{ch2}
\section*{Introduction}
La segmentation assure la décomposition d'une image en des parties sémantiquement homogènes~\cite{aubert2002}. Ce technique permet l'extraction de formes qui ont des caractéristiques communes par exemple: l'intensité de couleur, la texture, la forme géométrique, \textit{etc}~\cite{morel95}.
Mais, si la méthode de segmentation se base seulement sur les caractéristiques citées au-dessus, les résultats de l'extraction de la zone d'intérêt seront incohérents et incorrects. \\
En effet, Si nous voulons trouver une structure spécifique dans un tomogramme, pour plus de fiabilité, l'utilisation des méthodes de segmentation classiques ne sera pas suffisante. \\
En imagerie biologique, il y a deux défis~\cite{bio1}. Premièrement, la difficulté de la segmentation augmente lors de la phase de préparation de l'échantillon biologique et l'acquisition. Par exemple, la connectivité entre les régions d'une structure biologique peut être erronée vu que le contraste du tomogramme est inconstant et varie d'un angle à une autre. Ce qui fait des fissures peut apparaître sur l'échantillon~\cite{cryo_artefacts}.\\
Deuxièmement, la segmentation d'une image biologique exige l'identification d'objets multiples d'une image. Ces objets ont des formes hétérogènes et différentes aux formes connues telles que la forme circulaire, rectangulaire, \textit{etc}. Ce qui fait l'utilisation des modèles de formes géométriques est insuffisant. Aussi bien que les méthodes classiques de segmentation ne seront pas suffisantes pour segmenter le tomogramme.\\
Troisièmement, la configuration des caractéristiques de l'objet à extraire varie d'un type à un autre tel que la morphologie, l'intensité de couleur, la texture, \textit{etc}.

\section{Historique}
Au milieu des années 50, des systèmes ont été développés pour automatiser la classification de cellules exfoliées sur le frottis, dans le but de fournir le dépistage du cancer du col de l'utérus~\cite{morel95}. Ces systèmes sont basés sur le seuillage en utilisant un échantillon unidimensionnel (1D)~\cite{histcellseg1}.\\
en 1960, l'apparition de la première méthode automatisée de traitement d'images bidimensionnelles (2D) pour le comptage différentiel des globules blancs (leucocytes). Cette méthode est basée sur des mesures colorimétrie et morphologique~\cite{histcellseg2}.\\
Au milieu des années 70, les systèmes commerciaux qui assurent ce test clinique courant ont été mis en vente. Ces derniers contiennent plusieurs circuits informatiques qui assure la parallélisation des t\^{a}ches : l'analyse de l'image précédente de la cellule, tout en saisissant l'image actuelle et en même parallèle la localisation de la cellule suivante dans l'échantillon~\cite{histcellseg3}.\\
à cette époque, les premiers microscopes assistés par un ordinateur ont été mis en vente pour le traçage et l'analyse morphologique des cellules neuronales~\cite{histcellseg4}.\\
Dans les années 1980, le progrès des systèmes de microscope confocal (microscope optique) a ouvert la porte pour l'analyse d'images tridimensionnelles (3D). Mais, les ordinateurs sont devenus suffisamment puissants pour gérer les données 3D que dans les années 1990, aussi bien que des données 2D complexes comme en histopathologie~\cite{histcellseg5}. A partir de cette époque, les communautés de vision par ordinateur et de traitements d'images ont commencé à relever des défis~\cite{histcellseg5}.\\
Durant les dernières décennies, certaines recherches ont abordé ce sujet. La croissance des travaux qui ont été abordés est exponentielle. Plus que la moitié de la plupart des articles ont été apparus après l'année 2000. Les méthodes d'analyse d'images cellulaires publiées assurent le comptage de cellule, la déduction des types de cellule et des formes,~\textit{etc}~\cite{histcellseg6}.

\section{Approches de segmentation biologiques}
\begin{figure}[!htbp]
  \centering
\includegraphics[scale=0.7]{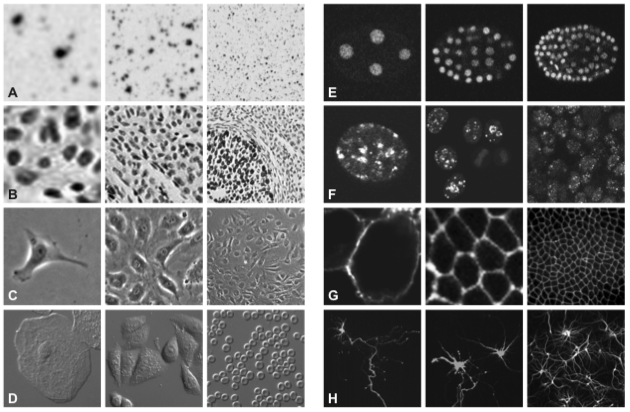} \\
\caption{Variété de types, densité et outils d'acquisition de cellules~\cite{cellseg}}\label{fig1cell} 
\end{figure}

La segmentation automatique des images biologiques, pour analyser les cellules, est généralement un problème difficile à résoudre.  à cause de la variété des microscopes, des types de cellules, des densités de cellules et la complexité des données à analyser (\textit{cf. figure}~\ref{fig1cell}). 

Comme la montre la figure~\ref{fig1cell}, la densité de cellules augmente du gauche à droite et les microscopes utilisés pour l'acquisition sont différents.  Microscopie en fond clair pour les images (A-B), microscopie en contraste pour l'image (C), microscopie en contraste d'interférence différentielle (D), microscopie en fluorescence (E-H). La majorité de cellules a une forme sphérique et le reste a une forme différente.  

Finalement, en examinant les travaux de recherche accomplis sur ce sujet depuis 1960. Nous constatons que la majorité des méthodes de segmentation cellulaire se base sur les approches suivantes :
 \renewcommand{\labelitemi}{$\blacksquare$}

\begin{itemize}
  \item Seuillage 
  \item Détection de caractéristiques
  \item  Filtrage morphologique
  \item  Accumulation de régions
  \item  Modèle déformable
\end{itemize}

\subsection{Seuillage}
L'une des approches les plus utilisées en segmentation cellulaire est le seuillage~\cite{cellseg13}. Généralement, l'intensité des cellules est différente par rapport à l'arrière-plan globalement ou localement (\textit{cf. figure}~\ref{fig2cell}). Pour le premier cas, l'utilisation d'un seuil fixe suffira. Pour le deuxième cas, il faut utiliser un seuillage adaptatif.  \\
à la suite d'une analyse statique, globalement et localement, des valeurs d'intensité globale ou local à partir de l'histogramme de l'image,  la sélection des seuils sera automatiquement effectuée.  \\
Comme la montre, la figure~\ref{fig2cell} 2A, les cellules ont une intensité lumineuse et sont parfaitement séparées de l'arrière-plan. Ce type d'image sera facilement segmenté en utilisant l'approche de seuillage.
Mais malheureusement en pratique, l'utilisation de seuillage seulement, produit des résultats médiocres lors de la segmentation cellulaire. Pour cette raison que la plupart des méthodes de segmentation appliquent le seuillage comme première étape et non l'unique étape du processus de segmentation.

\subsection{Détection de caractéristiques}
Au lieu de segmenter l'image par leur intensité, les cellules peuvent être segmentées en fonction de l'intensité de leurs caractéristiques qui sont facilement déduites à l'aide d'un filtrage linéaire.

Par exemple, avec un faible grossissement, les cellules ressemblent à des particules compactes et peuvent être déduites à l'aide d'un détecteur de taches tel que le filtre gaussien ou laplacien de Gaussienne (\ref{fig2cell}~.B). Dans le cas d'un fort grossissement, les cellules apparaissent plus larges, mais si leurs formes sont invariantes, un modèle de filtre dédié pourrait être dérivé de l'image.

 Comme la méthode de seuillage, l'utilisation de cette méthode seulement ne précise pas des résultats exacts. Ce filtre produit des indicateurs utiles pour des étapes ultérieures de détection du contour de la cellule.

\subsection{Filtrage morphologique}
Un autre type de filtrage connu est le filtrage morphologie provenant du domaine de la morphologie mathématique.
Les opérateurs tels que l'érosion, la dilatation, l'ouverture et la fermeture permettent d'examiner et de manipuler les propriétés géométriques et topologiques des objets dans une image. Ils sont fréquemment utilisés en liaison avec la segmentation cellulaire. 
Nous pouvons construire facilement des filtres en faisant appel à ces opérateurs spécifiés au-dessus.
Il faut distinguer entre la morphologie binaire et la morphologie des niveaux de gris~\cite{cellseg13}. Le premier est utilisé comme une étape de post-traitement pour améliorer les résultats obtenus par la segmentation (Figure~\ref {fig2cell}~.A), tandis que le deuxième est utilisé comme étant un prétraitement pour améliorer ou supprimer des structures d'image spécifiques pour la segmentation (Figure~\ref {fig2cell}~.C).

\subsection{Accumulation de régions}
Une autre approche de segmentation cellulaire appelée accumulation de régions. à partir des points de départ sélectionnés dans l'image,  la méthode forme des régions étiquetées en reliant de manière itérative les points sélectionnés par des points connectés.

La mise en oeuvre la plus simple de cette méthode est la fusion de région, qui fonctionne par couche de voisinage de points connectés et, appliquée directement à l'image.

 Des schémas hiérarchiques de scission et fusion, fonctionnant par couche de résolution et utilise un prédicat d'uniformité, sont généralement utilisés. Un autre exemple est la méthode de ligne de partage des eaux  (\textit{watershed})~\cite{cellseg13}, qui est la principale approche de segmentation issue de la morphologie mathématique, qui fonctionne par couche d'intensité et nécessite une image ayant un contour foncé (amplitude du gradient) (Figure~\ref {fig2cell}~.C).
 Bien que l'approche par accumulation de région soit de loin la plus populaire, la méthode \textit{watershed}  est connu par la production d'une sur-segmentation, et nécessite généralement un traitement supplémentaire.

\subsection{Modèle déformable}
\begin{figure}[!htbp]
  \centering
\includegraphics[scale=0.7]{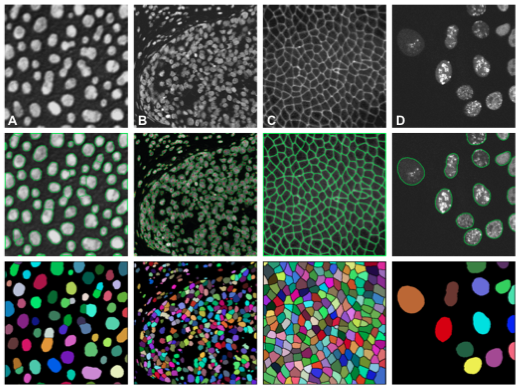} \\
\caption{Variété de types, densité et outils d'acquisition de cellules~\cite{cellseg}}\label{fig2cell} 
\end{figure}

La dernière approche de segmentation cellulaire est celle du modèle déformable. Cette approche consiste à ajuster les données de l'image par un modèle déformable (\ref{fig2cell}~.D).
 Les modèles déformables peuvent être formulés explicitement, sous forme d'un contour actif paramétrique (2D) ou de surface (3D), ou implicitement, comme le niveau zéro d'une fonction avec une dimension (nD) supérieure à la dimension de l'image à segmenter~\cite{cellseg16}.
Ce modèle est évolué de manière itérative afin de minimiser l'énergie fonctionnelle prédéfinie, qui se base sur les images ou les formes.

\section*{Conclusion}
Dans cette section, nous venons de décrire quelques méthodes de segmentation fréquemment utilisées dans le domaine de la biologie cellulaire. Tel que les méthodes de seuillage, détection de caractéristiques, filtrage morphologique, accumulation de région et modèle déformable. Par la suite, nous avons comparé les méthodes de segmentation biologique. Dans le chapitre qui suit, nous allons décrire notre approche.
\chapter{Méthode de segmentation semi automatique d'extraction de ribosomes} \label{ch3} %
\section*{Introduction}
\begin{figure}[!htbp]
  \centering
\includegraphics[scale=0.5]{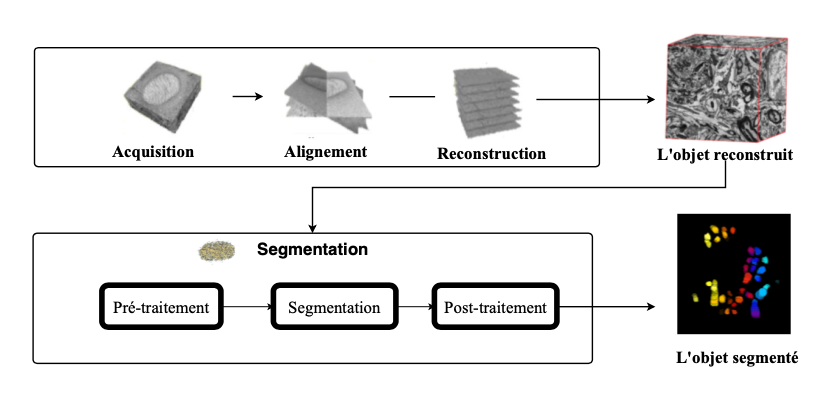} \\
\caption{Approche proposée}\label{fig1ch3} 
\end{figure}

Généralement, en cryotomographie, les images révèlent un problème au niveau du contraste. Pour cette raison, avant de passer à la phase de segmentation, une amélioration de qualité de contraste de l'image reconstruite doit être effectuée. \\
Il existe de nombreuses méthodes de segmentation, l'utilisation d'une seule méthode ne suffira pas pour obtenir de bons résultats et les résultats obtenus se diffèrent en changeant la nature de cellule à trouver et le type d'image acquise. Donc la complexité du choix de méthodes de segmentation est toujours élevée.  
 \\ Bien évidemment, après l'évaluation des résultats par un expert. Il est possible d'améliorer la solution et d'effectuer un post-traitement pour minimiser les erreurs et affiner les résultats.  \\ 
Comme c'est déjà mentionné dans les chapitres précédents, la reconstruction tomographie passe par 4 étapes, l'acquisition, l'alignement, la reconstruction et enfin la segmentation. Notre travail se focalise sur l'amélioration de phase de segmentation. En effet, nous allons segmenter les coupes 2D prises à partir de l'objet 3D reconstruit.  \\ 
Dans ce chapitre, nous allons décrire notre approche qui aura 3 étapes (\textit{cf. figure}~\ref{fig1ch3}). La première étape est un prétraitement qui a pour but d'améliorer le contraste des coupes 2D de l'objet reconstruit. La deuxième étape est la segmentation. à cette étape, nous utilisons plusieurs méthodes dont nous allons les décrire en détail. Et l'étape finale est un post-traitement pour corriger les erreurs et effectuer un deuxième alignement pour affiner les résultats.

\section{Pré-traitement}
Généralement, à la formation de l'image, lorsqu'une onde d'électrons traverse l'échantillon, deux changements s'effectuera, le premier est au niveau d'amplitude, le deuxième au niveau de l'onde d'électrons. En effet, l'apparition du contraste est toujours liée au choix de la phase d'onde et le niveau d'amplitude.

Nous pouvons définir le contraste par le quotient suivant qui est la différence d'intensité entre une région ayant un contraste élevé et son arrière-plan~\cite{metwc} :

 \begin{equation}
  C=\frac{I-I_{arriere\_plan}}{I_{arriere\_plan}}
\end{equation}

\begin{figure}[!htbp]
  \centering
  \includegraphics[width=0.5\linewidth]{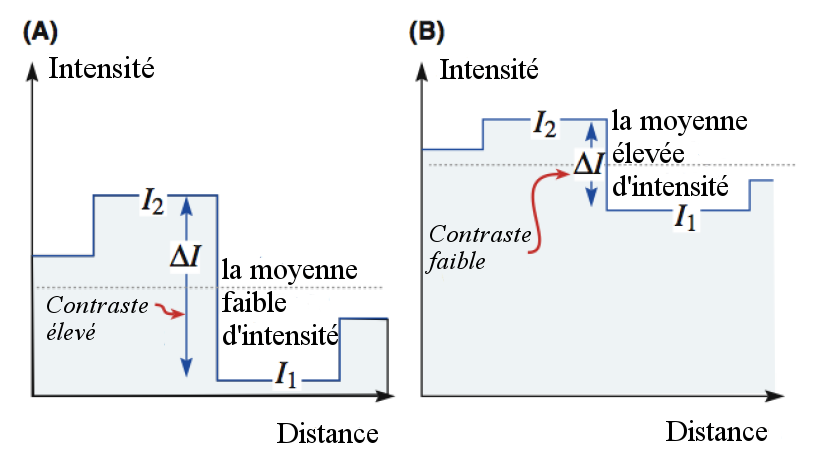}
  \caption{La différence entre le contraste et l'intensité. ($I_1$ et $I_2$) illustrent les niveaux d'intensité et ($\Delta I$) présente le contraste.}\label{contraste}
\end{figure}

La différence entre l'intensité et le contraste est que la première liée à la densité d'électrons qui peuvent salir l'écran ou le détecteur et l'intensité peut être claire ou foncée, par contre le contraste peut être élève ou faible (\textit{cf. figure}~\ref{contraste}).

\begin{figure}[!htbp]
  \centering
\includegraphics[scale=0.35]{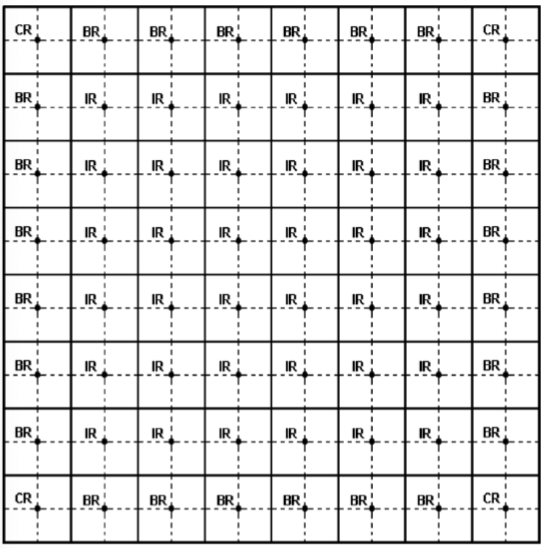} \\
\caption{Structure de régions avec une image de taille 512 x 512 divisé en 64 régions~\cite{clahe}}\label{square} 
\end{figure}

Concernant notre approche, nous avons utilisé la méthode CLAHE (\textit{Contrast limited adaptive histogram equalization}) comme étant un prétraitement pour améliorer le contraste des coupes 2D de l'objet 3D reconstruit.  
Dans plusieurs travaux, l'utilisation de la méthode CLAHE pour les images médicales a montré des résultats intéressants. 
Cette méthode est basée sur le fait de diviser l'image en des régions équivalentes. Par exemple, pour une image de taille 512x512, pour obtenir de bons résultats, généralement la division de régions est effectuée comme suit. Diviser l'image par 8 pour chacune direction. Il y'aura 6 directions donc le nombre de régions sera 64. Un autre exemple, mais cette fois pour montrer une partition de l'image et non sa totalité. Dans la figure~\ref{square}, la partition de l'image carrée de taille (512 x 512) contient 3 groupes de régions, le premier (1) groupe CR (\textit{corner regions}), le deuxième (2) est BR (\textit{boarder regions}) et le dernier (3) IR (\textit{inner regions}) respectivement 4 régions appartenant au premier groupe (1), 24 pour le deuxième(2), et enfin 36 pour le dernier (3). \\

Le processus de l'amélioration de contraste en utilisant la méthode CLAHE contient deux étapes principales: le calcul de la fonction de transfert du contraste et la combinaison de résultats~\cite{clahe}.

  Premièrement, concernant la première étape, la méthode CLAHE traite les groupes de régions (CR, BR et IR) séparément au lieu de l'image entière, en calculant la fonction de transfert de contraste de chacune. Par la suite, pour chaque groupe, le contraste sera amélioré, séparément, jusqu'à ce que l'histogramme du groupe amélioré se ressemble à l'histogramme spécifié lors de la distribution. 
  
Le calcul de l'histogramme s'effectue en calculant, pour chaque groupe de régions, le nombre de pixels appartenant à chaque niveau de gris. A la suite de ce calcul, l'histogramme de chaque groupe de régions sera généré. 
L'égalisation de l'histogramme est obtenue en utilisant la fonction CDF (cumulative distribution function). Si le nombre de pixels et le nombre des niveaux de gris, pour chaque groupe de régions, sont respectivement M et N, et si ($h_{i,j}(n)$ pour n = 0,1,2,...,N-1) est l'histogramme de la région (i,j) donc une estimation de la fonction CDF, mise à l'échelle correctement avec N-1 niveaux de gris, est comme suit :     
   \begin{equation}\label{clahe1}
  f_{i,j}(n) = \frac{(N-1)}{M} \cdot \sum_{k=0}^{n} h_{i,j}(k); 
   n= 1,2,3...,N-1
  \end{equation}
Cette fonction est utilisée pour rendre la fonction de densité de niveaux de gris uniforme. Comme c'est déjà mentionné, cette procédure est celle de l'égalisation de l'histogramme. \\
Pour chaque groupe de régions, pour chaque histogramme calculé, le problème affronté dans cette approche est la sursaturation de l'image qui est due à un maximum élevé dans l'histogramme ,surtout dans les zones homogènes, il y'aura toujours un nombre important de pixels situés dans la même plage de niveaux de gris.  \\
  Pour cette raison, il faut préciser une limite. L'une des possibilités est d'utiliser \textit{ClipLimit} $\beta$ pour ajuster tous les histogrammes, car sans fixer cette limite, la technique d'égalisation adaptative de l'histogramme pourrait produire des résultats parfois pires que l'image originale. Donc, la pente maximale de~\ref{clahe1} est limitée à une pente maximale fixée à l'avance.  
  \begin{equation}\label{clahe2}
\beta = \frac{M}{N} (1+\frac{\alpha}{100}(s_{max}-1))
\end{equation}

La modification de l'histogramme initial est basée sur le \textit{ClipLimit} $\beta$ qui désigne la limite que nous devrons le prendre en considération, en calculant le nombre de pixels appartenant à chaque niveau de gris. 
Si le nombre de pixels, appartenant à un niveau de gris quelconque, dépasse la limite fixée. Les pixels qui seront comptés au-delà de cette limite seront uniformément distribués à d'autres niveaux de gris qui n'ont pas encore dépassé cette limite. Cette redistribution, pour chaque histogramme, peut nécessiter plusieurs itérations jusqu'à ce qu'aucun niveau de gris ne dépasse la limite $\beta$. L'algorithme de la redistribution itérative de l'histogramme est illustré dans l'algorithme~\ref{algo1}. Pour chaque groupe de régions, nous obtenons l'histogramme ajusté en utilisant la fonction~\ref{clahe1}.
\begin{algorithm}[htb]
\caption{Redistribution de l'histogramme dans la méthode CLAHE~\cite{clahe}} \label{algo1}
Excess = 0 \newline
For n = 0,1,2,...,N-1\newline
If {$h(n) > \beta$}, Then \newline
$Excess \gets Excess + h(n) - \beta$ \newline
$h(n) \gets \beta$ \newline
End If \newline
End For \newline
m=Excess/N \newline
For n=0,1,2,...,N-1\newline
If {$h(n) < \beta - m$}, Then \newline
$h(n) \gets h(n) + m$ \newline
$Excess \gets Excess - m$\newline
Else If {$h(n) < \beta$}, Then \newline
$Excess \gets Excess - \beta + h(n)$ \newline
$h(n) \gets \beta$ \newline
End If \newline
End For \newline
While {$Excess > 0$} \newline
For n=0,1,2,...,N-1\newline
If Excess > 0 Then \newline
If {$h(n) < \beta$}, Then \newline
$h(n) \gets h(n) +1$ \newline
$Excess \gets Excess - 1$ \newline
End If\newline
End If\newline
End For\newline
End For\newline
End While
\end{algorithm}

\begin{figure}[!htbp]
  \centering
\includegraphics[scale=0.5]{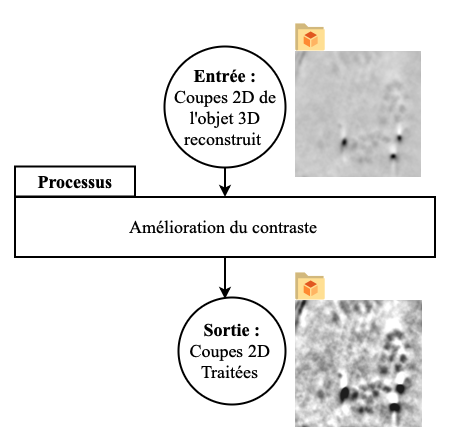} \\
\caption{Pré-traitement : amélioration du contraste}\label{fig2ch3} 
\end{figure}

 Deuxièmement, en utilisant une interpolation bilinéaire, tous les groupes de régions (CR, BR et IR) qui ont été améliorées seront unifiés. Dans le but d'éliminer les contours artificiels. \newpage
La figure~\ref{fig2ch3} illustre l'une des coupes 2D de l'objet reconstruit qui a été amélioré en utilisant la méthode CLAHE.

\section{Segmentation}
\begin{figure}[!htbp]
  \centering
\includegraphics[scale=0.7]{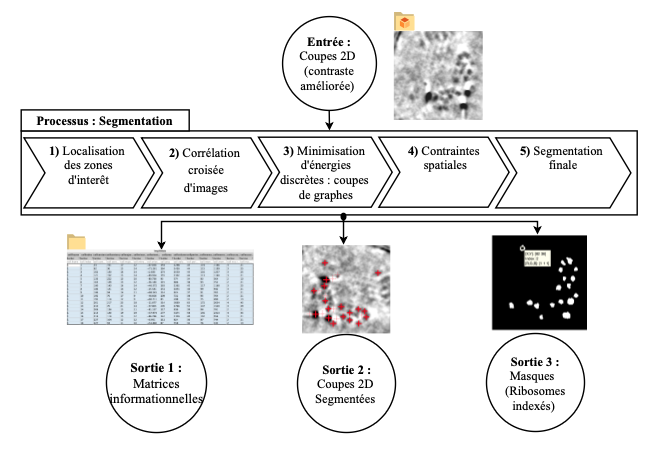} \\
\caption{Segmentation}\label{fig3ch3} 
\end{figure}

L'approche de segmentation cellulaire utilisée est basée sur la structure de la membrane. Dans cette section, nous allons nous référer à cette approche par l'abréviation MPCS \textit{(membrane pattern-based cell segmentation)}.  \\
Les étapes de l'approche MPCS sont les suivants  (\textit{cf. figure}~\ref{fig3ch3}) : 
\begin{enumerate}
\item Spécification de paramètres biologiques.
\item Localisation des zones d'interêt.
\item Corrélation croisée d'images.
\item Minimisation d'énergie discrètes : coupes de graphes.
\item Contraintes spatiales.
\item Segmentation finale.
\end{enumerate}

Nous allons décrire en bref l'approche MPCS  (\textit{cf. figure}~\ref{fig3ch3}) avant de détailler chaque étape mentionnée au-dessous.Tout d'abord en se basant sur l'une des coupes 2D de l'objet reconstruit. Nous spécifions quelques paramètres biologiques. Ensuite, nous détectons les points potentiels dans les cellules (\textit{seeds}). Chaque point d'intérêt sera traité individuellement. Faisons appel à la théorie des graphes, en combinant des opérations de corrélation croisée et corrélation des directions par coupes de graphe, un contour optimal pour chaque cellule sera détecté. La corrélation croisée est une technique qui mesure la similarité entre deux signaux. Dans notre cas, cette technique est utilisée pour décoder les informations liées au modèle de membrane. Ces informations seront ajoutées dans un graphe et en faisant appelle à des contraintes spatiales définies, nous utilisons les coupes de graphe pour extraire chaque cellule de son arrière-plan. Comme c'est déjà mentionné au-dessous, chaque point sera traité séparément. Donc tous les résultats de segmentation obtenus seront groupés pour obtenir le résultat de segmentation finale de l'image.

\subsection{Spécification de paramètres biologiques}
\begin{figure}[!htbp]
  \centering
\includegraphics[scale=0.55]{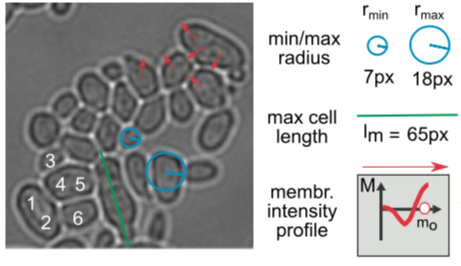} \\
\caption{Spécification de paramètres cellulaires~\cite{biosol}}\label{a} 
\end{figure}

Les paramètres biologiques de MPCS peuvent être définis de manière interactive par l'interface utilisateur graphique (GUI) de CellX~\cite{cellx}.
 Par exemple, la figure~\ref{a} illustre l'exemple de cellule de levure en herbe de morphologie aberrante, nous estimons d'abord le rayon maximal et rayon minimal  ($r_{min}$, $r_{max}$; en bleu) (sachant qu'il est possible de spécifier plusieurs rayons minimaux et maximaux). Par la suite, nous spécifions la longueur de la plus longue cellule. ($l_{m}$ ; ligne verte). Nous estimons le modèle de la membrane cellulaire (vecteur d'intensité M, représenté dans la courbe rouge) en faisant la moyenne d'intensités d'un ensemble de profils de membranes (tracé de l'intérieur à l'extérieur de la cellule, illustré par des flèches rouges). 

\subsection{Localisation des zones d'intérêt}
 
\begin{figure}[!htbp]
  \centering
\includegraphics[scale=0.55]{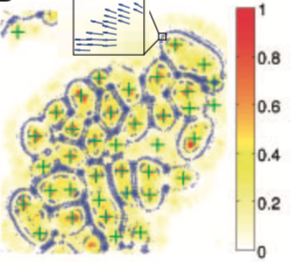} \\
\caption{Application du transformé de Hough et déduction des \textit{seeds}~\cite{biosol} }\label{b} 
\end{figure}

Dans la deuxième étape, nous identifions les zones d'intérêt de cellules. Dans la plupart des cas, en appliquant quelques opérations traditionnelles de traitement d'images nous pouvons conclure les points d'intérêt. Dans notre approche, nous allons appliquer le transformé de Hough (à partir du gradient)~\cite{gbha} sur les coupes 2D de l'objet reconstruit.\\
$L=W \times H$ , $W={1,',w}$, $H={1,',h}$ , o\`{u} w est la largeur de l'image et h sa longueur (mesurées en nombre de pixels). \\
Nous calculons le gradient de l'image (flèches bleues (\textit{cf. figure}~\ref{b})) et   nous vérifions pour chaque pixel, s'il est dans la même direction que le vecteur gradient (à une distance comprise entre $r_{min}$ et $r_{max}$), nous calculons une valeur approximative de la magnitude du gradient. Les résultats de tous les pixels de l'image seront stockés dans une matrice d'accumulation (région codée par couleur  (\textit{cf. figure}~\ref{b})).
Nous détectons les régions de l'image ayant des maxima locaux et nous utilisons leurs centres comme un point d'intérêt \textit{seed} conclu (croix vertes (\textit{cf. figure}~\ref{b})). Pour chaque \textit{seed} trouvé : s = ($x_{s}$, $y_{s}$) , o\`{u} $x_{s} \in W $ et $y_{s} \in H$, nous calculons également la distance radiale $r_{c} \in$ [$r_{min, rmax}$]  o\`{u} s est le centre et  $r_{c}$ est le rayon.

\subsection{Corrélation croisée d'images}
Séparément, pour chaque \textit{seed}, une fenêtre carrée sera générée dont le centre est le \textit{seed} (recadrage avec une largeur de 2$l_{m}$ + 1); et un nouveau point de départ à s = (l$_{m}$, l$_{m}$). 
Nous utilisons d'abord l'algorithme de Bresenham~\cite{crosscorr1} pour déterminer les pixels situé dans chaque segment. En effet, le point de départ du segment sera le \textit{seed} et le point final sera chaque pixel appartenant aux contours de I. Ensuite, en utilisant la corrélation croisée~\cite{crosscorr2} , nous calculons ensuite les profils d'intensité de ces segments linéaires (rayons), dans le but de déduire leurs similarités avec le modèle de membrane M. 
 
Les valeurs de corrélation croisée seront élevées au niveau des régions qui sont compatibles avec le modèle de membrane et faibles dans les régions qui ne seront pas similaires au modèle. 
 
\begin{figure}[!htbp]
  \centering
\includegraphics[scale=0.55]{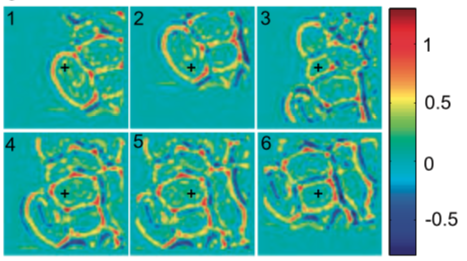} \\
\caption{Corrélation croisée d'images~\cite{biosol}}\label{c} 
\end{figure}

 Les valeurs de corrélation croisée de tous les rayons nous permettent de générer une image de corrélation croisée (\textit{cf. figure}~\ref{c}) que nous allons l'utiliser par la suite  pour formuler le problème de traçage des contours de cellules sur un graphe. 
 
 En effet, nous nous référons à CC (p) en tant que le résultat de corrélation croisée attribué au pixel p = (x$_{p}$, y$_{p}$) dans l'ensemble de pixels $P = X \times X$, X = {1, ... , 2l$_{m}$ + 1} de I.
 
\subsection{Minimisation d'énergie discrètes : coupes de graphes}
Pour chaque recadrage de l'image I, nous utilisons les coupes de graphe pour extraire la région d'intérêt (la cellule) de son arrière-plan, aussi bien que les cellules voisines. Nous attribuons une étiquette de type binaire (cellule ou arrière-plan)$ A_{p}$ pour chaque pixel $p \in P$.
Nous combinons ces affectations dans un vecteur A ayant une dimension 1 x $\left | P \right |$ (d'o\`{u} $\left | P \right |$ = $10^{4}$ , pour les images de cellule avec $l_{m}$ = 50 pixels).
Tout d'abord, nous définissons un graphe orienté $G = (V, E, w)$ avec l'ensemble de sommets $V = P$. Nous définissons bien évidemment le voisinage du rayon r, d'un pixel $p \in P$, par la fonction \[N(p,r) = \left \{ q \in P\left \| p-q \right \|_{2} \leqslant r \right \}\] pour définir le contour comme étant \[E = \bigcup_{p \in P } \bigcup_{q \in N (p,\sqrt{2})}^{p\not\equiv q} (p,q)\]

 Ceci lie chaque pixel dans I à ses voisins dans toutes les directions (vertical horizontale et diagonal). Les contours sont associées à un poids défini par la fonction $w : E \rightarrow \mathbb{R}$. à partir lequel nous définissons la fonction d'énergie : 
\[EN(A)=\sum_{(p,q)\in E} w((p,q)).id(A_{p}\neq A_{q})\]

o\`{u} w (p, q) mesure le co\^{u}t d'affecter deux pixels voisins à différentes partitions (informations sur le controur) et la fonction id (.) vaut 1 si la condition à l'intérieur des parenthèses est vraie si non elle vaut 0. Nous déterminons le vecteur optimal A qui minimise EN (A) en déterminant la coupe minimale sur le graphe d'image défini précédemment. L'algorithme min-cut / max-flow~\cite{mincoupegraphe1} fournit le résultat optimal en temps polynomial. Selon~\cite{mincoupegraphe2}, si nous définissons les poids des arêtes comme suit : 
\[w((p,q)) =\frac {\delta^{2} .\left | \epsilon _{pq} \right |^{2}.\Delta \phi .det(D(p))}{2.\left [\epsilon_{pq}^{T}.D(p).\epsilon _{pq} \right ]^{\frac{3}{2}}}\]

alors la \textit{min-cut} correspond à la géodésique de la cellule traitée.

 \noindent$\delta$, désigne la longueur de c\^{o}tés des pixels quadrilatère convexes ayant quatre c\^{o}tés de même longueur (égale à 1 dans notre cas); \\
 q$_{pq}$, le vecteur reliant les sommets de graphes p et q; \\
 $\Delta \phi$, la différence d'orientation angulaire des vecteurs caractéristiques de la grille (égale à $\pi$ / 4  pour notre système à 8 quartiers); \\
 det (.), l'opération déterminante; et en pixel p, D (p) désigne le tenseur métrique qui est défini comme suit : 
\[r_{c}/2, C=N(s,r_{c}/2) \subset P\]

\[D(p)=g(p).M_{id} +(1-g(p)).u(p).u(p)^{T}\]

o\`{u} $M_{id}$ est la matrice d'identité, u (p) est un vecteur unitaire qui a la même direction que  le gradient de l'image au niveau du pixel p et g (p) est la fonction scalaire qui parcours la magnitude à partir les informations de contours fournit dans le graphe (les poids). Nous utilisons la fonction exponentielle et incluons les valeurs de corrélation croisée~\cite{mincoupegraphe2} sous la forme : \[g(p)=exp(-10^{CC(p)})\]
 
\subsection{Contraintes spatiales}

\begin{figure}[!htbp]
  \centering
\includegraphics[scale=0.55]{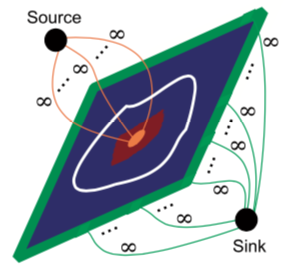} \\
\caption{Contraintes spatiales~\cite{biosol}}\label{d} 
\end{figure}

Pour éviter les solutions triviales et limiter le contour de la cellule, nous imposons des contraintes spatiales lors de la génération de coupe de graphe, en transformant le problème en un problème de flot maximal sur un réseau (multisource) (\textit{cf. figure}~\ref{d}).
 Plus précisément, nous définissons une région circulaire autour du \textit{seed} de rayon $r_{c}/2$, \[C = N(s, r_{c} /2) \subset P\] qui doit appartenir à la cellule (région orange (\textit{cf. figure}~\ref{d})) ainsi qu'au pixel défini dans la partie rognée de l'image \[B= \left \{ p \in P \left |N(p,\sqrt{2}) \right | < 8 \right \} \subset P\] (région verte (\textit{cf. figure}~\ref{d})).  
 En utilisant des pondérations pour connecter les pixels de C et B aux sommets sources et aux puits donnés, nous limitons les coupes de graphes à la région située entre C et B (contour blanc (\textit{cf. figure}~\ref{d})). 

\newpage
\subsection{Segmentation finale}
\begin{figure*}[!htbp]
\begin{multicols}{2}
\includegraphics[scale=0.4]{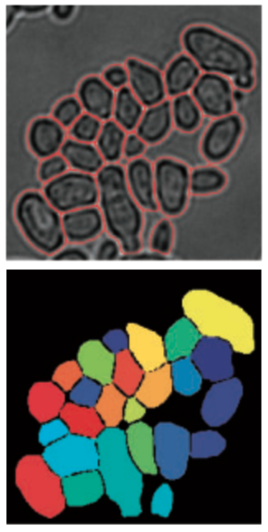} \par
\includegraphics[scale=0.4]{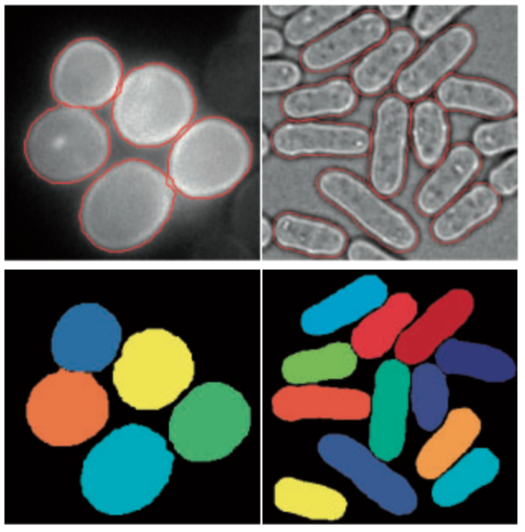} \par
    \end{multicols}
\caption{Segmentation finale}\label{ef}
\end{figure*}

Rappelons que, dans cette approche, dans les sous-sections précédentes, chaque \textit{seed} a été traité séparément et à cette étape, à partir des résultats obtenus séparément, nous devrons conclure l'image finale segmentée. Tout d'abord, nous allons éliminer chaque  \textit{seed} ayant une valeur aberrante, en prenant compte sa morphologie et la valeur de sa corrélation croisée. Ensuite, nous allons fusionner les  \label{seeds}  ayant des régions identiques dans l'image. Enfin, nous allons corriger les petits chevauchements de segmentations cellulaires en affectant les pixels à la cellule la plus proche. Ces étapes assurent que chaque pixel de l'image est attribué à l'arrière-plan ou à l'une des cellules. La figure~\ref{ef} montrent quelques exemples d'images segmentées par cette approche.

\begin{figure}[!htbp]
  \centering
\includegraphics[scale=0.65]{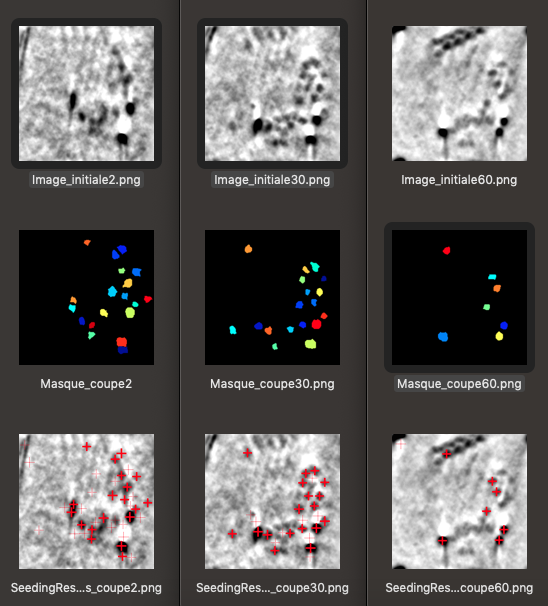} \\
\caption{Segmentation des coupes 2D de ribosomes }\label{ribseg} 
\end{figure}

En effet, nous allons segmenter les coupes 2D de ribosomes en utilisant cette approche. La figure~\ref{ribseg},  illustre quelques coupes 2D segmentées.

En observant les coupes 2D de la figure~\ref{ribseg}, les ribosomes segmentés sont incohérents et non alignés. Donc, l'utilisation de cette approche de segmentation ne suffira pas pour obtenir de bons résultats. 
\newpage
\section{Post-Traitement}
\begin{figure}[!htbp]
  \centering
\includegraphics[scale=0.6]{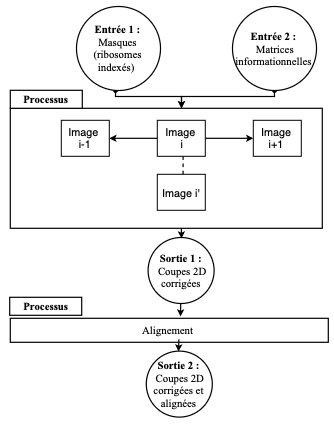} \\
\caption{Minimisation du taux d'erreur et alignement }\label{fig4ch3} 
\end{figure}

Vu l'incohérence des résultats obtenus durant l'étape précédente de segmentation, nous avons ajouté une phase de post-traitement pour améliorer les résultats de segmentation. Dans ce chapitre, nous allons décrire notre méthode d'amélioration et dans le chapitre qui suit nous allons illustrer en détail les résultats obtenus. \\
Ce traitement aura deux étapes, la première est la minimisation de taux d'erreur en comparant les coupes 2D segmentées entre eux et généré des nouvelles coupes cohérentes. La deuxième étape est d'aligner à nouveau ces nouvelles coupes (\textit{cf. figure}~\ref{fig4ch3})).\\

\newpage
\subsection{Minimisation du taux d'erreur}
\begin{figure}[!htbp]
  \centering
\includegraphics[scale=0.6]{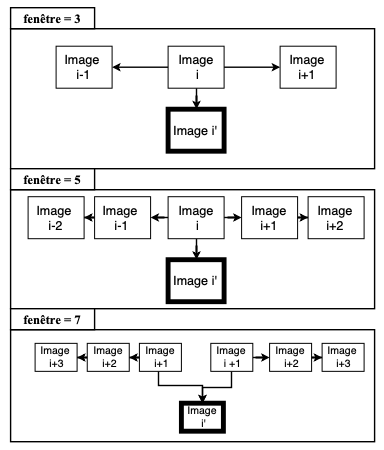} \\
\caption{Minimisation du taux d'erreur  : taille de la fenêtre }\label{window} 
\end{figure}

La comparaison des coupes 2D de ribosome pour obtenir des coupes 2D cohérentes dépend de la taille de fenêtre. La taille de la fenêtre est égale à 3 ou 5 ou 7 (\textit{cf. figure}~\ref{window}), respectivement, compare l'image courante, par l'image précédente et suivante, par deux images précédentes et suivantes, par trois images précédentes et suivantes (\textit{cf. figure}~\ref{fig4ch3}).

\begin{figure}[!htbp]
  \centering
\includegraphics[scale=0.4]{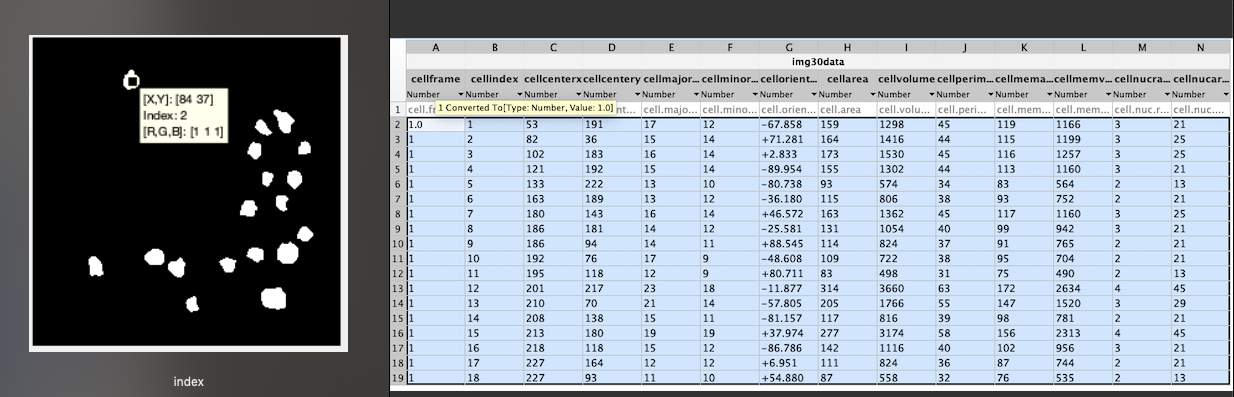} \\
\caption{Matrice informationnelle de cellules }\label{corrcell} 
\end{figure}

\newpage
Le processus de minimisation d'erreur que nous allons décrire sera appliquée d'une manière itérative. Chaque coupe 2D segmentées sera traitée séparément, en comparant ces cellules par rapport aux cellules par rapport à la coupe précédente et suivante. 
Sachant que la méthode de segmentation fournit une matrice informationnelle de cellules (\textit{cf. figure}~\ref{corrcell})). 

\newpage
\noindent $im$ désigne la coupe 2D.  \\
($N$) est le nombre totale des coupes 2D. \\
$im\left [ i \right ]$ est la i ème coupe. \\
Chaque cellule de la coupe 2D (image i) est définie par un index $i_{c}$.  \\
x et y sont les coordonnées du centre de la cellule ayant un index $i_{c}$.  \\
($A$) désigne l'aire de la cellule ayant un index $i_{c}$.  \\
Et enfin ($P$) qui sera le périmètre de la cellule ayant un index $i_{c}$. \\
$im_{new } \left [ i \right ]$ est la nouvelle coupe générée ($i$ ème).
     
Pour le cas o\`{u} f=3 : \\
 Pour chaque image $im\left [ i \right ]$, le parcours se fait comme suit, tel que $i \in \left [ 2 , N-1 \right ] $). Nous comparons les valeurs du centre (x,y), l'aire A et le périmètre P de chaque cellule de l'image $im\left [ i \right ]$  avec l'image précédente $im\left [ i-1 \right ]$  et suivante $im\left [ i+1 \right ]$ . En fixant 3 seuils pour le centre, l'aire et le périmètre, si la différence entre eux ne dépasse pas les seuils fixés, nous garderons la cellule lors de la génération de la nouvelle coupe 2D cohérente $im_{new}$. Sinon la cellule sera éliminée.
     \\
     
Pour le cas o\`{u} f=5 :\\
 Pour chaque image $im\left [ i \right ]$, le parcours se fait comme suit, tel que $i \in \left [ 3, N-2 \right ] $ ). Nous comparons les valeurs du centre (x,y), l'aire A et le périmètre P de chaque cellule de l'image $im\left [ i \right ]$  avec ces deux prédécesseurs  $im\left [ i-1 \right ]$ et $im\left [ i-2 \right ]$ aussi bien que ces deux successeurs  $im\left [ i+1 \right ]$ et $im\left [ i+2 \right ]$  . En fixant 3 seuils pour le centre, l'aire et le périmètre, si la différence entre eux ne dépasse pas les seuils fixés, nous garderons la cellule lors de la génération de la nouvelle coupe 2D cohérente $im_{new}$. Sinon la cellule sera éliminée.  \\
 
Pour le cas o\`{u} f=7 :\\
 Pour chaque image $im\left [ i \right ]$, le parcours se fait comme suit, tel que $i \in \left [ 4 , N-3 \right ]$  ). Nous comparons les valeurs du centre (x,y), l'aire A et le périmètre P de chaque cellule de l'image $im\left [ i \right ]$  avec ces trois prédécesseurs  $im\left [ i-1 \right ]$, $im\left [ i-2 \right ]$ et $im\left [ i+3 \right ]$, aussi bien que ces trois successeurs  $im\left [ i+1 \right ]$, $im\left [ i+2 \right ]$ et $im\left [ i+3 \right ]$  . En fixant 3 seuils pour le centre, l'aire et le périmètre, si la différence entre eux ne dépasse pas les seuils fixés, nous garderons la cellule lors de la génération de la nouvelle coupe 2D cohérente $im_{new}$. Sinon la cellule sera éliminée.  \\

\subsection{Alignement}
Nous avons utilisé la méthode d'alignement par corrélation croisée pour aligner les images de projections 2D segmentées que nous avons améliorées.

Cette méthode définit une fonction discrète $f(m, n)$ de corrélation croisée entre deux coupes 2D voisines $I_1$ et $I_2$~\cite{Frank1992}:
\begin{equation}\label{crosss}
  f(m,n)=\frac{1}{MN}\sum_{i=0}^{M-1}\sum_{j=0}^{N-1}I_1(i,j)I_2(i+m,j+n)
\end{equation}
o\`{u} $M$ et $N$ illustrent le nombre de pixels dans les directions verticales et horizontales des coupes 2D.
Le processus d'alignement consiste à trouver un déplacement $(m_0,n_0)$, de fa\c{c}on que la fonction $f(m_0,n_0)$ abouti sa valeur maximale. 
Quand nous obtenons les estimations de déplacement relatif à un nombre suffisant de séries d'inclinaison, il est possible d'aligner les coupes 2D en appliquant des translations successives avec l'inverse des valeurs de décalage calculées.
 
Il est possible d'optimiser le temps de calcul de la corrélation en faisant appel à la transformée de Fourier et son inverse~\cite{Frank1992}:
\begin{equation}\label{croos_freq}
  f(m,n)=TF2^{-1}[TF2(I_1(i,j)).TF2(I_2(i+m,j+n))]
\end{equation}

En effet, pour résumer, le processus d'alignement par corrélation croisée sera comme suit :  \\
\begin{enumerate}
\item Transformée de Fourier
\item Corrélation croisée
\item Estimation des paramètres de transformation
\item Detections des pics
\item Inverse de la transformée de Fourier
\end{enumerate}
\newpage
\section*{Conclusion}
Dans ce chapitre, nous avons décrit les différentes étapes de notre approche. Commen\c{c}ant par le post-traitement qui assure l'amélioration du contraste des coupes 2D, passant par la segmentation, et enfin le post-traitement, qui traite les coupes 2D segmentées pour qu'elles soient cohérentes et alignées. Dans le chapitre suivant, nous allons illustrer en détail les résultats obtenus et les évaluer.


\chapter{Etudes expérimentales} \label{ch4} 
\section*{Introduction}
Comme nous avons mentionné dans le chapitre précédent, notre approche assure la segmentation des coupes 2D du tomogramme. Nous avons amélioré tout d'abord le contraste, par la suite nous avons segmenté les coupes 2D et enfin nous avons traité les résultats obtenus pour rendre les cellules segmentées cohérentes et alignées.
Dans ce dernier chapitre, nous allons évaluer cette approche. Nos tests sont faits avec  61 orientations avec un pas d'inclinaisons entre 60$^{\circ}$ et 60$^{\circ}$ qui est égal à 2. 

\section{L'impact du pré-traitement}
\begin{figure}[!htbp]
  \centering
\includegraphics[scale=0.45]{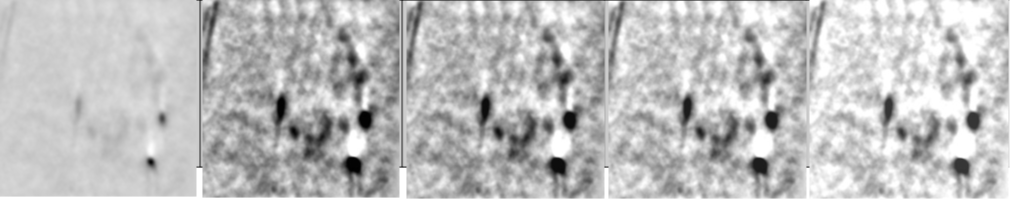} \\
\caption{Amélioration du contraste}\label{contrast} 
\end{figure}

Dans le chapitre précédent, nous avons mentionné que la première étape de notre approche est l'amélioration de contraste en utilisant la méthode CLAHE. Comme la montre la figure~\ref{contrast}, l'image à gauche est l'image initiale et les 4 images qui suivent sont les coupes 2D modifiées. Les quatre exemples illustrés dans cette figure ont des niveaux de contraste différents. Le but de modifier la valeur des paramètres plusieurs fois est de déduire la meilleure configuration.

\begin{figure}[!htbp]
  \centering
\includegraphics[scale=0.45]{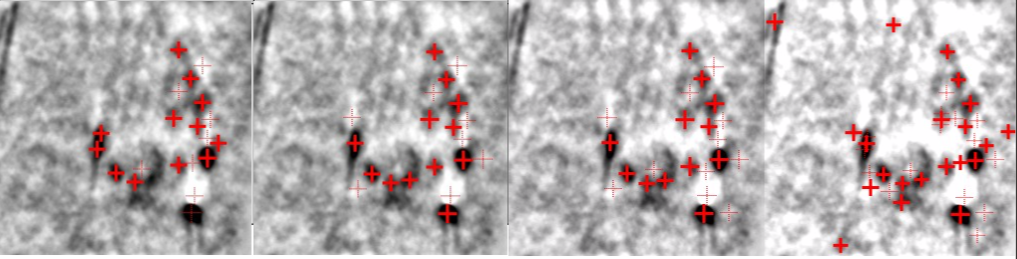} \\
\caption{Impact du pré-traitement à la segmentation}\label{figcont1} 
\end{figure}

\begin{figure}[!htbp]
  \centering
\includegraphics[scale=0.45]{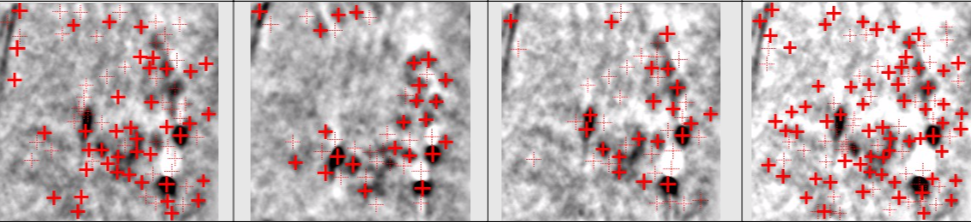} \\
\caption{Impact du pré-traitement à la segmentation}\label{figcont2} 
\end{figure}

Mais comme l'impact de l'amélioration ne sera visible qu'une fois, la segmentation est effectuée, la décision de la meilleure configuration a été prise juste après la phase de la segmentation. Pour justifier notre constatation, nous avons choisi deux coupes prises à partir d'un angle différent. La première figure~\ref{figcont1} montre un résultat de segmentation identique pour toutes les images. Et la deuxième~\ref{figcont2} illustre l'exemple donné par  la plupart des coupes qui montre des résultats de segmentation totalement différente. Nous avons évalué les 61 coupes 2D, mais nous avons constaté qu'une seule configuration parmi les quatre que nous avons vérifiés a montré la stabilité de ces résultats pour les 61 coupes.

\section{L'impact du post-traitement}
Le processus du post-traitement s'effectue en deux phases, la première s'agit d'une correction d'erreur en comparant les différentes coupes 2D segmentées à fin de déduire les cellules cohérentes, par la suite, la deuxième s'agit d'effectuer un alignement à nouveau.

\begin{figure}[!htbp]
  \centering
\includegraphics[scale=0.75]{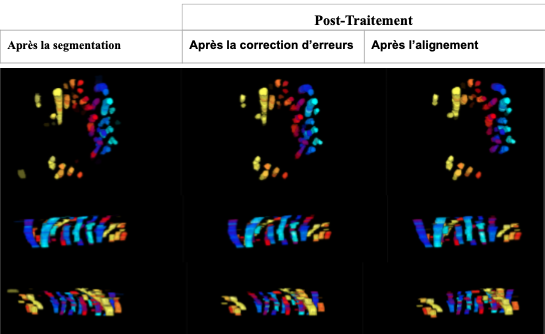} \\
\caption{L'impact du post traitement sur la segmentation avec une fenêtre de taille 3}\label{win3t} 
\end{figure}

\begin{figure}[!htbp]
  \centering
\includegraphics[scale=0.75]{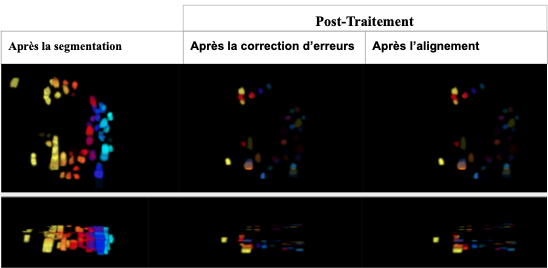} \\
\caption{L'impact du post traitement sur la segmentation avec une fenêtre de taille 5}\label{win5t} 
\end{figure}

\begin{figure}[!htbp]
  \centering
\includegraphics[scale=0.75]{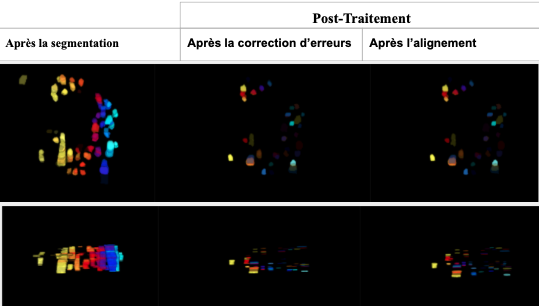} \\
\caption{L'impact du post traitement sur la segmentation avec une fenêtre de taille 7}\label{win7t} 
\end{figure}

Nous avons varié la taille de la fenêtre comme c'est déjà mentionné dans le chapitre précédent. La taille de la fenêtre pour les figures~\ref{win3t}, \ref{win5t} et~\ref{win7t} est respectivement 3, 5 et~7. 

Visuellement, les résultats obtenus avec une fenêtre de taille 3 sont meilleurs que les autres. Et nous allons affirmer notre constatation par la suite en calculant la précision et le rappel.

\newpage

\section{Evaluation}
L'évaluation de l'approche nécessite la définition d'un critère qui permet d'estimer les résultats. Nous avons donc utiliser les deux mesure rappel et précision vu que l'aspect qualitatif est plus important dans notre approche. 
Le rappel est égal au nombre d'individus correctement détectés par rapport aux individus qui doivent être détectés. Dans notre exemple, l'individu désigne le ribosome. Donc, cette dernière mesure la capacité de l'approche à l'extraction de ribosomes.

Nous allons calculer le rappel en utilisant la formule suivante :

\begin{equation}\label{rappel}
  \textmd{rappel}=\frac{\textmd{Nombre\ d'objets\ correctement\ identifiés}}{\textmd{Nombre\ d'objets\ pertinents\ dans\ l'image}}
\end{equation}

Revenant maintenant à la deuxième mesure, la précision illustre le nombre d'individus correctement identifiés par rapport au nombre d'individus total. Cette dernière assure l'évaluation de la capacité du système à la restitution d'individus pertinents. 

Nous allons calculer la précision en utilisant la formule suivante :
\begin{equation}\label{precision}
  \textmd{précision}=\frac{\textmd{Nombre\ d'objets\ correctement\ identifiés}}{\textmd{Nombre\ d'objets total}}
\end{equation}

\begin{figure}[!htbp]
  \centering
\includegraphics[scale=0.6]{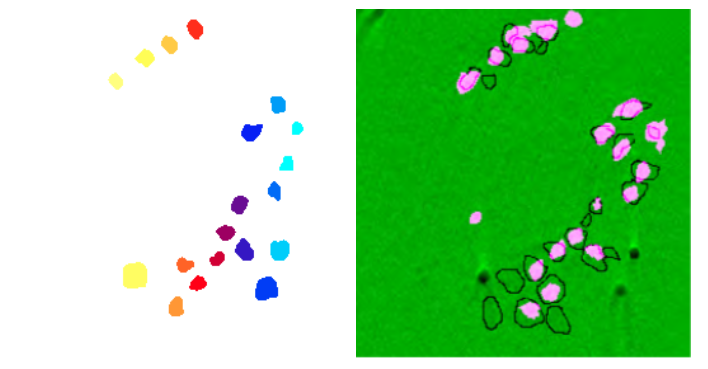} \\
  \caption{Comparaison entre les résultats obtenus et les ribosomes identifiés par l'expert.}\label{figc51}
\end{figure}

En analysant visuellement la partie gauche de la figure~\ref{figc51} nous constaterons que les résultats obtenus lors de l'extraction des ribosomes sont fiables. En comparant les ribosomes identifiés par l'expert et notre résultat nous confirmons notre constatation précédente. La figure montre que la plupart des ribosomes ont été parfaitement identifiés.  

Les valeurs  de rappel et précision (\textit{cf. equation}~\ref{r} et~\ref{p}) en se référant à la figure~\ref{figc51} précisée par l'expert, valide qualité des résultats que notre approche fournit.
 
\begin{equation}\label{r}
  \textmd{rappel}=  65 \%  
\end{equation}

\begin{equation}\label{p}
    \textmd{précision}= 78 \% 
\end{equation}

\section*{Conclusion}
Nous avons illustré tous les résultats obtenus durant les différentes phases de notre approche. Nous avons bien évidemment évalué la première phase d'amélioration de contraste et montré son impact sur les résultats de segmentation. Nous avons aussi évalué la méthode de segmentation aussi bien que la phase de post-traitement et les améliorations qu'elles assurent.
\Conclusion{Conclusion et Perspectives} \label{Conclusion}

\PARstart{D}{ans} le domaine de la biologie cellulaire 3D, la qualité des images utilisées dans la reconstruction et les angles d'acquisition manquants rendent l'extraction des particules d'intérêt assez difficile. La complexité de la segmentation augmente en fonction de la dimension de l'image à segmenter. Un problème de segmentation d'une image 2D ne sera pas traité de la même manière qu'un problème de segmentation 3D.  \newline
Notre travail consiste à proposer une approche de segmentation du tomogramme. Deux phases de préparation et filtrage accompagnent notre approche proposée : une phase de prétraitement précède la phase de segmentation de notre approche de segmentation semi-automatique hybride (2D-3D). Aussi bien qu'elle suit phase de post-traitement.
Dans la première phase, nous avons essayé d'améliorer les coupes 2D provenant du tomogramme. Bien évidemment, nous avons utilisé la méthode CLAHE. Par la suite, nous avons segmenté les coupes 2D améliorées en utilisant une adaptation d'un algorithme d'extraction des structures membrane. Avant de finir, nous avons traité les coupes 2D segmentées d'une fa\c{c}on itérative pour améliorer leurs cohérences. Enfin nous avons aligné de nouveau ses coupes segmentées améliorées en utilisant la méthode d'alignement par corrélation croisée afin d'assurer la continuité volumique de chaque objet extrait..\newline
Afin d'évaluer la pertinence de notre approche, nous avons appliqué les étapes précédemment définies sur les images.\newline
En effet, les expérimentations réalisées ont montré que l'extraction à l'aide de techniques proposés a amélioré les résultats, l'utilisation des relations entre les coupes 2D, pour avoir une information sur la totalité de volume, aide à mieux sélectionner les segments d'intérêt et les artefacts. Aussi bien que les phases de pré- et post- traitements proposées ont amélioré les résultats obtenus. Toutes ces remarques sont illustrées par les bons résultats que nous avons obtenus en termes de rappel et de précision de notre classifieur.

\newpage

Sachant que la solution que nous avons proposée est une amélioration d'un autre travail de recherche \cite{tomog3d}. Nous pouvons bien évidemment améliorer notre travail, ainsi en ajoutant d'autres modules, par exemple, vu que la première étape est la plus importante, plus que la qualité d'image est améliorée plus que l'ambigu\"{i}té sera réduite dans les phases qui suivent. Nous pouvons proposer une méthode de filtrage spécifique, destinée à être utilisée par les images biologiques qui représente les ribosomes ou des cellules ayant les mêmes caractéristiques. Cette méthode sera basée sur l'amélioration des points faibles de l'image biologique tels que le bruit, le contraste et la morphologie cellulaire.

\nocite{ref1,ref2,ref3,ref4,ref5,ref6,ref7,ref8,ref9,ref10,ref11,ref12,ref13,ref14,ref15,ref16,ref17,ref18,ref19,ref20}


\cleardoublepage \addcontentsline{toc}{chapter}{\bibname}%
\bibliographystyle{alpha-fr}
\bibliography{memoire}
\cleardoublepage

\pagenumbering{Roman}
\setcounter{page}{1}

\end{document}